\documentclass[11pt]{article}
\pdfoutput=1 
\usepackage{jheppub}

\usepackage{CJKutf8} 
\usepackage{mathtools,slashed,mathrsfs}
\usepackage[caption=false]{subfig}
\usepackage{dcolumn}
\usepackage{makecell}
\usepackage{multirow}
\usepackage{tabularx}
\usepackage{booktabs}
\usepackage{orcidlink}
\usepackage{bm}
\usepackage{amsmath}
\usepackage{amsthm}
\usepackage[normalem]{ulem}
\usepackage{comment}
\usepackage{setspace}
\usepackage{xcolor}
\usepackage{cancel}
\usepackage{enumerate}
\usepackage{enumitem}
\usepackage{siunitx}
\usepackage{float}
\usepackage{hyperref}
\usepackage{cancel}
\usepackage{tensor}

\allowdisplaybreaks

\newcommand{\pwrap}[1]{\mathopen{}#1\mathclose{}\mathord{\vphantom{#1}}} 
\newcommand{\p}[1]{\pwrap{\left(#1\right)}}
\newcommand{\bq}[1]{\pwrap{\left[#1\right]}}
\newcommand{\Bq}[1]{\pwrap{\left\{#1\right\}}}
\newcommand{\bk}[1]{\pwrap{\left\langle#1\right\rangle}}

\newcommand{\abs}[1]{\left|#1\right|}
\newcommand{\eval}[1]{\left.#1\right|}

\newcommand{\lbar}[1]{\,\overline{\!#1}\vphantom{#1}}

\newcommand{\Cite}[1]{ref.~\cite{#1}}
\newcommand{\fig}[1]{figure~\ref{#1}}

\newcommand{\tab}[1]{table~\ref{#1}}
\newcommand{\eq}[1]{eq.~(\ref{#1})}
\newcommand{\Eq}[1]{Eq.~(\ref{#1})}
\newcommand{\eqs}[2]{eqs.~(\ref{#1}) and (\ref{#2})}
\newcommand{\Sec}[1]{section~\ref{#1}}

\newcommand{\peq}{\phantom{{}={}}}

\newcommand{\n}{\nonumber\\}

\newcommand{\dd}[2][]{\mathop{\mathrm{d}^{#1}#2}\nolimits}

\DeclareSIUnit\parsec{pc}

\definecolor{plotred}{HTML}{DC143C}
\definecolor{plotblue}{HTML}{0018A8}

\newcommand{\Rc}{\mathcal{R}} 
\newcommand{\phipt}{\delta\phi} 
\newcommand{\phicl}{\phi_0} 
\newcommand{\dphicl}{\dot\phi_0}

\newcommand{\eff}{\mathrm{eff}}
\newcommand{\Ld}{\mathcal{L}} 

\newcommand{\Ne}{\mathcal{N}_e}
\newcommand{\fNL}{f_{\mathrm{NL}}} 

\newcommand{\LambdafD}{\Lambda_{\mathrm{5D}}} 
\newcommand{\LambdaIR}{\Lambda_{\mathrm{IR}}} 
\newcommand{\Lambdaeff}{\Lambda_{\eff}} 

\newcommand{\gf}[1]{\mathcal{#1}} 

\newcommand{\ctime}{\eta} 
\newcommand{\Mpl}{M_{\mathrm{pl}}} 
\newcommand{\Mc}{M_{\mathrm{C}}} 
\newcommand{\Fosc}{F_{\text{osc}}} 

\newcommand{\Ps}{\mathcal{P}} 
\newcommand{\Op}{\mathcal{O}} 

\newcommand{\Z}{\mathbb{Z}} 

\newcommand{\vb}[1]{\mathbf{#1}} 
\newcommand{\order}{\mathcal{O}}

\newcommand{\cc}{\mathrm{c.c.}} 

\title{\centering Extra-dimensional Origins of Chemical Potentials\\at the Cosmological Collider}

\date{\today}
\author[a]{Raman Sundrum \orcidlink{0009-0004-7537-5357},}
\author[a]{Zhaohui Xu \orcidlink{0009-0001-0473-1121} }

\affiliation[a]{Maryland Center for Fundamental Physics, Department of Physics, University of Maryland, College Park, MD 20742, USA}

\emailAdd{raman@umd.edu}
\emailAdd{zhxu1226@umd.edu}

\abstract{We study the realization of the chemical potential mechanism in cosmological collider physics in a robust class of inflationary models arising from multiple higher-dimensional gauge fields. The rolling inflaton background corresponds to an electric field in the extra dimension in which charged particles are produced analogously to the Schwinger mechanism, while neutral particles can be produced through non-minimal interactions. We show that particles heavier than the inflationary Hubble scale can be created without Boltzmann suppression in both cases. In particular, charged Kaluza-Klein excitations can be created through minimal gauge interactions. We construct realistic models along these lines consistent with both theoretical and experimental constraints.}

\begin{document}

\maketitle

\section{Introduction}

\label{sec:intro}

During high-scale inflation, extremely heavy particles can be produced by the time-dependent background and ``recorded" in primordial fluctuations through mechanisms collectively known as \emph{cosmological collider physics} \cite{Chen:2009zp,Chen:2012ge,Noumi:2012vr,Assassi:2012zq,Arkani-Hamed:2015bza,Lee:2016vti}.
With dramatic improvements in ongoing and upcoming Cosmic Microwave Background (CMB) experiments \cite{CMB-S4:2016ple,Sohn:2019rlq,NASAPICO:2019thw}, Large Scale Structure (LSS) surveys \cite{Alvarez:2014vva,SPHEREx:2014bgr,Camera:2014bwa,MoradinezhadDizgah:2017szk,MoradinezhadDizgah:2018ssw,Kogai:2020vzz,Euclid:2024yrr}
and more futuristic \qty{21}{cm} tomography \cite{Loeb:2003ya,Cooray:2006km,Munoz:2015eqa,Meerburg:2016zdz}, there is a real possibility of observing such heavy particles far beyond the reach of terrestrial colliders.
In the minimal mechanism of cosmological collider physics, particles are created by the Hubble expansion, with distinctive oscillatory signatures of observable strength only appearing for masses in a narrow window of \(\order(H)\) \cite{Arkani-Hamed:2015bza,Lee:2016vti}. Fortunately, there is a powerful extension by the \emph{chemical potential mechanism} \cite{Barnaby:2011vw,Chen:2018xck,Adshead:2018oaa,Wang:2019gbi,Wang:2020ioa,Bodas:2020yho,Tong:2022cdz,Bodas:2024hih,Bodas:2025wuk}, where heavy particles are produced by coupling to the rolling inflaton background \(\phi_0(t)\), potentially allowing a mass reach up to \(\order\Bigl(\sqrt{\dphicl} \approx 60H\Bigr)\) \cite{Planck:2018jri}.

Despite its exciting prospects for phenomenology, the chemical potential mechanism raises significant conceptual issues for Effective Field Theory (EFT) control. As we will detail below, these issues are similar in spirit to the well-known problem of realizing a trans-Planckian inflaton field range required in high-scale inflation \cite{Linde:1983gd,Lyth:1996im,Arkani-Hamed:2003xts}. In this paper, we will show how an attractive class of solutions to the trans-Planckian problem within higher-dimensional gauge theory automatically gives rise to chemical potentials and resolves these issues.

In more detail, the chemical potential mechanism is based on a dimension-5 interaction,
\begin{align}\label{eq:4D-chem-potential}
    \delta\Ld &= -\frac{1}{\Lambda} \nabla_\mu\phi J^\mu,
\end{align}
where \(\phi\) is the inflaton and \(J^\mu\) is some current bilinear in the heavy fields. This interaction allows particles with masses up to the chemical potential \(\lambda = \dphicl/\Lambda\) to be produced.
For a mass reach \(\order(10H)\), say, the scale of non-renormalizability \(\Lambda\le\order(400H) \ll \Mpl\).
Meanwhile, assuming that the rolling inflaton background lasts for \(\Ne\) \(e\)-foldings, the field distance the inflaton travels has to be much greater than \(\Lambda\):
\begin{align}\label{eq:trans-Lambda}
    \frac{\Delta\phi}{\Lambda} \simeq \frac{\dphicl \Delta t}{\Lambda} = \frac{\dphicl}{\Lambda}\cdot\frac{\Ne}{H} = \Ne \p{\frac{\lambda}{H}} \gg 1,
\end{align}
i.e. the naïve EFT expansion in \(\phi/\Lambda\) clearly breaks down for such a small \(\Lambda\).
At first sight, this can be evaded if the inflaton is predominantly derivatively coupled as a \emph{Pseudo-Nambu-Goldstone Boson} (PNGB), so that while \(\Delta\phi\) is large, the EFT expansion is in derivatives of \(\phi\) which can still be small. The simplest example is an ``axion'', a PNGB associated to spontaneously broken global \(\mathrm{U}(1)\). However, if \(\Lambda\gtrsim f\), the PNGB decay constant, the problem still persists as now \eq{eq:trans-Lambda} implies \(\Delta\phi> 2\pi f\), inconsistent with the latter being the maximal field range. On the other hand, while it appears hard to achieve \(\Lambda\ll f\),  we will show later in this paper how this can be effectively realized.\footnote{There are other proposals to realize the approximate shift symmetry besides PNGB of internal symmetries, e.g. flat moduli in supersymmetry and brane inflation. As far as we know, none of these alternatives alone have established a fully satisfactory solution to the \(\eta\) problem, as have been discussed in \cite{Arkani-Hamed:2003wrq} for supersymmetry and reviewed in \cite{Kachru:2003sx} for brane inflation. We will nevertheless restrict the scope of this paper to PNGB as the origin of the shift symmetry.}

The trouble from large field distance is analogous to the trans-Planckian problem that preexists in high-scale inflation. According to the Lyth bound \cite{Lyth:1996im}:
\begin{align}
    \frac{\Delta\phi}{\Mpl} &\simeq \Ne \sqrt{2\epsilon_H},&
    \epsilon_H &\coloneqq \frac{\dphicl^2}{2\Mpl^2H^2} < 0.0044\ \text{\cite{Planck:2018jri}},
\end{align}
high-scale single-field inflation with \(\epsilon_H\approx0.004\) (\(H\approx\qty{6e13}{GeV}\)) requires the inflaton to have a flat potential over a trans-Planckian field range during inflation.
While one may attempt to protect the flatness of the potential by an approximate shift symmetry for a PNGB, near the Planck scale the requisite global symmetry and spontaneous symmetry breaking are not expected to survive quantum gravity violations \cite{Kallosh:1995hi}.

\begin{figure}
    \centering
    \raisebox{-0.4\totalheight}{\includegraphics{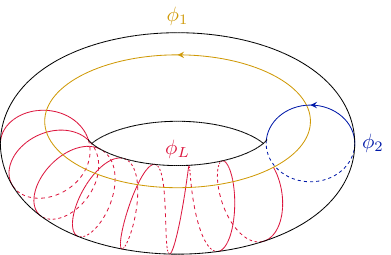}}\qquad
    \raisebox{-0.5\totalheight}{\includegraphics{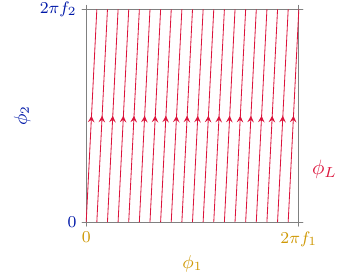}}
    \caption{The field configuration space of the bi-axion model \eq{eq:bi-axion-potential} with \(N=20\). The inflaton is identified as the light mode and its trajectory is shown in red. For sufficiently large \(N\), the length of this trajectory can be longer than \(\Mpl\) even with sub-Planckian \(f_1\) and \(f_2\)}
    \label{fig:bi-axion}
\end{figure}

In both cases, the need for a large inflaton field range in high-scale inflation raises concern over the validity of EFT. We will focus on one of the simplest ways of bringing this under EFT control, namely the \emph{aligned axion mechanism} \cite{Kim:2004rp,Choi:2014rja}. The inflaton field range is identified with an axion-like decay constant \(f_\eff\gtrsim\Mpl\), but emergent from an effective light degree of freedom inside the field space of \emph{multiple} axions with sub-Planckian decay constants. For example, with two axions, the field space is fundamentally a torus, on which the effective inflaton spirals \(N\gg1\) times in one of its periodical direction, as shown in \fig{fig:bi-axion}. In this case, \(f_\eff\) is given by
\begin{equation}
    f_\eff \simeq Nf,
\end{equation}
so that the fundamental decay constant \(f\) can be sub-Planckian and within EFT control. The same model simultaneously solves the large field range problem for the chemical potential in \eq{eq:trans-Lambda} as now \(\Delta\phi_2/f\) can be as large as \(N\).

Even with multiple axions, the fundamental decay constants \(f\) are still not far below \(\Mpl\), and the existence of global symmetries at such scales is still not robust against quantum gravity corrections. In the aligned axion mechanism, the spiral trajectory required to solve the trans-Planckian problem is enforced by a specific scalar potential and is thus sensitive to any new source of explicit symmetry breaking.
The simple solution is to have the global symmetries arise accidentally in the \emph{infrared}. The weakly-coupled option is when the PNGBs arise as extra-dimensional components of higher-dimensional gauge fields, in which case gauge invariance and higher-dimensional locality can protect against quantum gravity corrections above the compactification scale \cite{Arkani-Hamed:2003xts}.
In string theory, with more extra dimensions, one can also find candidates for the inflaton through compactification of higher-form gauge fields \cite{Svrcek:2006yi}.
Purely in 4D, PNGBs can emerge as composites of strongly-coupled dynamics, analogously to the pion of QCD. In those cases where the 4D dynamics has a holographic dual, the PNGBs are again extra-dimensional components of gauge fields in the higher-dimensional dual (for review and references, see, e.g., \Cite{Contino:2003ve}).

Interestingly, in such higher-dimensional gauge theoretic formulations of inflation, the rolling inflaton background, \(\dphicl\), corresponds to a higher-dimensional electric field. We will show that the chemical potential effects that produce heavy charged particles naturally arise in analogy to the \emph{Schwinger pair production mechanism} \cite{Sauter:1931zz,Heisenberg:1936nmg,Schwinger:1951nm}.
In this way, solving the trans-Planckian problem of high-scale inflation contains a powerful chemical potential mechanism as a corollary.
The extra-dimensional Schwinger pair production has been explored in \cite{Furuuchi:2015foh,Yamada:2024aca,Yamada:2025gvu}, though here we are choosing an alternative framework which is more promising for cosmological collider phenomenology.
Furthermore, we will also show that even neutral particles can be produced through non-minimal interactions with these gauge fields.
We will investigate the different chemical potential phenomena and estimate the viable mass-range of heavy particles and the signal strengths within EFT control.  In particular, we will show that extra-dimensional Kaluza-Klein excitations can be within the range of chemical potentials.

\section{Inflation in 4D EFT}

In this section, we introduce the notion of cosmological correlators and give a brief review of the chemical potential mechanism and the aligned-axion mechanism. In this paper, we adopt the mostly plus signature \((-,+,+,+)\) for the metric.

\subsection{Review of primordial fluctuations}

During slow-roll inflation, the spacetime is approximately described by the de Sitter metric: 
\begin{align}
    \dd{s}^2 &= -\dd{t}^2 +e^{2Ht} \dd{\vb{x}}^2\n
    &= \frac{-\dd{\ctime}^2 +\dd{\vb{x}}^2}{H^2\ctime^2},
    \label{eq:dS-metric}
\end{align}
where \(t\) is the proper time, \(H\) is the Hubble parameter during inflation and \(\ctime = -e^{-Ht}/H\) is the conformal time. Inflation is usually realized by a scalar field \(\phi\) called the ``inflaton", whose potential energy during inflation approximates a positive cosmological ``constant'' that drives the accelerated expansion of the universe. At classical level, the inflaton field is homogeneous and acts as the ``clock'' of the inflationary era. The classical inflaton field rolls down its potential and terminates inflation once the potential energy becomes subdominant.

At quantum level, the inflaton \(\phi\) must be treated as a quantum field that can be decomposed into a homogeneous classical background, \(\phicl\), plus a quantum fluctuation, \(\phipt\), that is responsible for the primordial fluctuations after inflation. In the spatially flat gauge, \(\phipt\) is related to the gauge-invariant comoving curvature perturbation, \(\Rc\), by \cite{Bardeen:1983qw}
\begin{equation}
    \Rc = -H \frac{\phipt}{\dphicl}
\end{equation}
with \(\dot x \coloneqq \dd{x}/\dd{t}\). Since the inflationary dynamics is invariant under spatial translation, it is convenient to work in spatial momentum basis where all momentum-space correlators contain a total-momentum \(\delta\)-function:
\begin{equation}
    \bk{\Rc_{\vb{k}_1}\Rc_{\vb{k}_2}\cdots\Rc_{\vb{k}_n}} = (2\pi)^3\delta(\vb{k}_1+\vb{k}_2+\cdots+\vb{k}_n)\cdot \bk{\Rc_{\vb{k}_1}\Rc_{\vb{k}_2}\cdots\Rc_{\vb{k}_n}}',
\end{equation}
where we use prime to denote correlators with the \(\delta\)-function stripped off.

Starting from the 2-point correlation function, current CMB observation favors a nearly scale-invariant primordial power spectrum given by
\begin{align}\label{eq:R-power-spectrum}
    \Ps_\Rc(k) \coloneqq \bk{\Rc_{\vb{k}} \Rc_{-\vb{k}}}' &= \frac{H^4}{2\dphicl^2 k^3} \p{\frac{k}{k_*}}^{n_s-1},& n_s &\approx 0.965\quad\p{k_* = \qty{0.05}{\mega\parsec^{-1}}}.
\end{align}
The fact that \(n_s\) is close to 1 implies that (1) the spacetime scale transformation:
\begin{equation}
    \ctime,\vb{x} \mapsto \lambda\ctime,\lambda\vb{x}
\end{equation}
is a good approximate symmetry, and (2) the inflaton is almost massless during inflation.
In the late-time approximation, \(\eta\to0\), to  reheating, (1) reduces to a purely spatial scale invariance, \(\vb{x}\mapsto\lambda\vb{x}\). As long as these two conditions hold, the power spectrum is restricted to be proportional to \(1/2k^3\). 

On the other hand, the 3-point correlation function \(\bk{\Rc_{\vb{k}_1}\Rc_{\vb{k}_2}\Rc_{\vb{k}_3}}'\), a.k.a. bispectrum, can contain more complicated momentum dependence even in the limit of scale invariance. The bispectrum is conventionally normalized relative to the power spectrum as \cite{Kumar:2017ecc}
\begin{align}\label{eq:3pt-F}
    F(k_1,k_2,k_3) &\coloneqq \frac{5}{6} \frac{\bk{\Rc_{\vb{k}_1}\Rc_{\vb{k}_2}\Rc_{\vb{k}_3}}'}{\Ps_\Rc(k_1) \Ps_\Rc(k_2) +\Ps_\Rc(k_2) \Ps_\Rc(k_3) +\Ps_\Rc(k_3) \Ps_\Rc(k_1)}\n
    &= -\frac{10}{3} \dphicl\cdot \frac{k_1^3 k_2^3 k_3^3}{k_1^3 +k_2^3 +k_3^3} B(k_1,k_2,k_3),
\end{align}
where \(B(k_1,k_2,k_3) = \bk{\phipt_{\vb{k}_1}\phipt_{\vb{k}_2}\phipt_{\vb{k}_3}}'\) is the 3-point function of inflaton fluctuations. CMB observation has also imposed constraints on the bispectrum, which are usually quoted in terms of \(\fNL\) defined by the equilateral configuration as
\begin{equation}
    \fNL \coloneqq F(k,k,k).
\end{equation}
The current bound on \(\fNL\) from CMB is around \(\order(10)\) for various analytic shapes of the bispectrum \cite{Planck:2019kim}, whereas the oscillatory cosmological collider signals are less constrained \cite{Sohn:2024xzd,Kumar:2026ogn,Kumar:2026dih}.
Meanwhile, ongoing and upcoming CMB and LSS experiments are expected to probe these signals at \(\fNL=\order(1)\) \cite{Sohn:2019rlq,Alvarez:2014vva,SPHEREx:2014bgr,Camera:2014bwa,MoradinezhadDizgah:2017szk,MoradinezhadDizgah:2018ssw,Kogai:2020vzz,Euclid:2024yrr,CMB-S4:2016ple,Sohn:2019rlq,NASAPICO:2019thw}, and more futuristic \qty{21}{cm} tomography can further bring this precision down to \(\fNL=\order(0.01)\) \cite{Loeb:2003ya,Cooray:2006km,Munoz:2015eqa,Meerburg:2016zdz}.

\subsection{Chemical potentials in phenomenological models}

\label{subsec:chem-pot-pheno}

In the minimal cosmological collider setup, the only source of energy for particle production is set by \(H\), 
with higher mass production being ``Boltzmann-suppressed''. Fortunately, the higher source of energy in the form of \(\dphicl \approx (60H)^2\) can be harnessed for particle production above \(H\) by ``chemical potential'' couplings to the inflaton: 
\begin{align}\label{eq:4D-chem-potential-2}
    \delta\Ld &= -\frac{1}{\Lambda} \nabla_\mu\phi J^\mu,
\end{align}
where \(J^\mu\) is a current bilinear in a heavy field. 
After plugging in the slow-roll approximated \(\phi \approx  \dphicl t\), this yields a chemical potential \(\lambda = \dphicl/\Lambda\) which can increase the mass reach up to \(\order(\lambda)\). A prerequisite for this is that the symmetry generated by \(\int\dd[3]{\vb{x}} J_0\) must be explicitly broken by the dynamics, otherwise the operator above can be eliminated by field redefinition.  

For the current \(J^\mu\), two classes of options have been proposed in the literature:
\begin{enumerate}
    \item \(\mathrm{U}(1)\) currents for complex scalars, spin-1/2 and spin-1 \cite{Chen:2018xck,Bodas:2020yho,Bodas:2024hih,Bodas:2025wuk}. For some complex scalar \(\chi\), this is
    \begin{equation}
        J^\mu = i\p{\bar\chi\nabla^\mu\chi -\chi\nabla^\mu\bar\chi}.
    \end{equation}
    It only becomes physical after adding  an explicit \(\mathrm{U}(1)\)-breaking term, e.g., a linear term in \(\chi\). The full Lagrangian is
    \begin{align}\label{eq:scalar-tree-model-chem-basis}
        \mathcal{L}_\chi &= -\abs{\nabla\chi}^2 -m^2 \abs{\chi}^2 -\frac{1}{\Lambda} \nabla_\mu\phi J^\mu -\frac{1}{\Lambda^2} \p{\nabla\phi}^2 \abs{\chi}^2
        -\alpha \chi -\bar{\alpha} \bar\chi.
    \end{align}
    For this class, there is another more transparent representation of the chemical potential. Take the scalar model \eq{eq:scalar-tree-model-chem-basis} as an example, one can perform a field redefinition \(\chi \mapsto \chi e^{i\phi/\Lambda}\) to the new frame as follow \cite{Bodas:2020yho}:
    \begin{align}\label{eq:scalar-tree-model-exp-basis}
        \mathcal{L}_\chi &= -\abs{\nabla\chi}^2 -m^2 \abs{\chi}^2 -\alpha \chi e^{i\phi/\Lambda} -\bar{\alpha} \bar\chi e^{-i\phi/\Lambda}\n
        &= -\abs{\nabla\chi}^2 -m^2 \abs{\chi}^2 -\alpha \chi e^{i\lambda t} e^{i\phipt/\Lambda} -\bar{\alpha} \bar\chi e^{-i\lambda t} e^{-i\phipt/\Lambda}.
    \end{align}
    Under this new frame, on-shell production of heavy particles is induced by explicit energy injections (or removal) \(=\lambda\) coming from the symmetry-breaking term.
    \item helicity-dependent currents for some massive vector boson \(B_\mu\) \cite{Wang:2020ioa}:
    \begin{align}\label{eq:vector-loop-model}
        J^\mu &= \frac{1}{4} \epsilon^{\mu\nu\rho\sigma} B_\nu G_{\rho\sigma} & \p{G_{\mu\nu}\coloneqq \partial_\mu B_\nu -\partial_\nu B_\mu}.
    \end{align}
    A generalization to spin-2 has also been discussed in \cite{Tong:2022cdz}. Unlike in the first class, the charge corresponding to \(J^\mu\) is not conserved at free field level (even in the massless limit), so no extra ``symmetry-breaking" term is needed for it to have physical effects.
\end{enumerate}

Now consider the cosmological collider signal from these models. In the scalar model \eq{eq:scalar-tree-model-chem-basis}, the cosmological collider signal in the bispectrum is produced by tree-level exchange of the heavy scalar. With \(\lambda>m\gg H\) and small \(\alpha\), it has a simple form given by \cite{Bodas:2020yho}
\begin{equation}
    F(k_1,k_2,k_3) \to \Fosc \p{\frac{2k_2}{k_1}}^{-3/2-i(\lambda-\mu)} +\text{c.c.},\qquad (k_2\approx k_3\gg k_1)
\end{equation}
where
\begin{align}\label{eq:f-arushi-model}
    \Fosc &\simeq \frac{5\pi}{6} \p{\frac{\alpha}{\Lambda}}^2 \frac{\lambda^{3/2} \mu^{-1/2}}{\lambda^2 -\mu^2} +\order\p{\alpha^4},&
    \begin{dcases}
        \lambda \coloneqq \frac{\dphicl}{\Lambda},\\
        \mu \coloneqq \sqrt{m^2 -\frac{9}{4}} \simeq m,
    \end{dcases}
\end{align}
and we have set \(H=1\) in these expressions.
To estimate the maximal signal size, we need to constrain the tadpole coefficient \(\alpha\). Note that the tree-level exchange of \(\chi\) also generates a correction to the power spectrum. According to \cite{Bodas:2025wuk}, for \(\lambda>\mu\), this correction is given by
\begin{align}\label{eq:power-spectrum-constraint}
    \delta\Ps(k) &\simeq \frac{4\pi\lambda}{\p{\lambda^2-\mu^2}^2} \p{\frac{\alpha}{\Lambda}}^2 \Ps_0(k),&
    \Ps_0(k) &\coloneqq \frac{H^4}{2\dphicl^2 k^3}.
\end{align}
In the perturbative region where \(\delta\Ps\ll \Ps_0\), we find
\begin{equation}\label{eq:f-arushi-model-2}
    \Fosc \simeq \frac{5}{24} \p{\frac{\delta\Ps}{\Ps_0}} \cdot \lambda^{1/2}\mu^{-1/2} \p{\lambda^2-\mu^2} \sim \p{\frac{\delta\Ps}{\Ps}} \lambda^2.
\end{equation}
For example, with \(\delta\Ps<0.1\Ps_0\), the maximal signal strength is \(\order\p{0.1-10^2}\) for \(\lambda\sim\mu\) from \(H\) to \(\order(60H)\). The non-perturbative region \(\delta\Ps\gtrsim\Ps\) has been studied in \cite{An:2017hlx,Huenupi:2026aqc,Huenupi:2026abj} and will not be considered in this paper.

For models where the heavy particles are produced at loop-level, the signals are typically much smaller and harder to estimate. Still, it has been found that in models of scalars and vectors, the loop-induced signal may exhibit exponential enhancement which can potentially overcome the loop suppression and lead to large cosmological collider signals. This is known to happen in the scalar model \eq{eq:scalar-tree-model-chem-basis} but with a quadratic \(\mathrm{U}(1)\)-breaking term \(\chi^2\) \cite{Bodas:2025wuk} and in the massive vector model with the helicity-dependent chemical potential \eq{eq:vector-loop-model} \cite{Wang:2019gbi,Wang:2020ioa}. In the latter case, the enhancement factor has a simple form \cite{Wang:2020ioa}:
\begin{align}\label{eq:helicity-exp-enhance}
    e^{6\pi (\lambda -\mu_B)},&&
    \mu_B \coloneqq \sqrt{m^2_B -\frac{1}{4}} \simeq m_B.
\end{align}
Therefore, even with loop suppression, a potentially observable signal can still be generated through this enhancement.

\subsection{Bi-axion model and chemical potentials}

\label{subsec:bi-axion}

In the theory of inflation with a single axion \(\phi\), the inflaton potential is generated from an explicit soft breaking of the global \(\mathrm{U}(1)\) symmetry, resulting in a potential \(V(\phi)\) which is \(2\pi f\)-periodic in \(\phi\). In the limit of small breaking, \(V(\phi)\) typically reduces to a single cosine model:
\begin{align}\label{eq:natural-inflation}
    V(\phi) &= V_0 \p{1 -\cos\frac{\phi}{f}}.
\end{align}
For this particular form of potential, known as \emph{natural inflation} \cite{Freese:1990rb,Adams:1992bn}, Planck 2018 results suggest that \cite{Planck:2018jri}
\begin{align}\label{eq:f_eff}
    f &\approx \frac{1}{\sqrt{-2\eta_V}} \Mpl \approx 10\Mpl,&
    \eta_V \coloneqq \Mpl^2 \frac{V_{\phi\phi}}{V} \approx -0.005
\end{align}
with \(\epsilon \approx 0.004\) for high-scale inflation. While Planck 2018 analysis ultimately disfavors this model at \(95\%\) CL, the tension can be addressed in multiple ways, the simplest of which will be including higher harmonics in the potential with modest tuning (see, e.g., \Cite{Czerny:2014wza,Kappl:2015esy}). On the contrary, the trans-Planckian value of \(f\) cannot be solved without significant tuning and thus presents a severe theoretical challenge as no global symmetry is expected to survive beyond \(\Mpl\) within a theory of quantum gravity.

Fortunately, the trans-Planckian problem can be addressed within an EFT of multiple axions through the aligned axion mechanism. In the bi-axion version, the two axions \(\phi_1,\phi_2\) have a special form of potential as follows \cite{Kim:2004rp,Choi:2014rja}:
\begin{equation}\label{eq:bi-axion-potential}
    V(\phi) = V_0 \p{1 -\cos\frac{\phi_1}{f_1}} +V_1 \bq{1 -\cos\p{-\frac{N\phi_1}{f_1} +\frac{\phi_2}{f_2}}}.
\end{equation}
With \(N\gg1\), we get two hierarchical mass eigenstates, where the light mode \(\phi_L\) is identified as the inflaton. Integrating out the heavy mode enforces the constraint
\begin{equation}\label{eq:bi-axion-traj}
    \frac{N\phi_1}{f_1} -\frac{\phi_2}{f_2} = 0.
\end{equation}
The effective potential for the light mode \(\phi_L\simeq \phi_2\) is thus given by
\begin{align}\label{eq:natural-inflation-effective}
    V(\phi_L) &= V_0 \p{1 -\cos\frac{\phi_L}{f_{\eff}}},&
    f_{\eff} \simeq Nf_2.
\end{align}
Consequently, one can have a trans-Planckian value for \(f_{\eff}\) with sub-Planckian values for the underlying \(f_1,f_2\) with a large \(N\). The CMB bound \eq{eq:f_eff} requires \(N\gg 10\). One can also generalize this to include higher harmonics for perfect realism, but we will not pursue it here for simplicity,

Now consider interactions between the inflaton and other fields. Since the inflaton \(\phi_L\) is almost aligned with the \(\phi_2\) direction, interactions involving \(\phi_2\) dominate over those of \(\phi_1\) and we will only consider them.
According to \Sec{subsec:chem-pot-pheno} and the periodicity of \(\phi_2\), the following two types of interactions are allowed and of interest:
\begin{align}\label{eq:2-types}
    \Op e^{iQ\phi_2/f_2},\qquad \frac{1}{\Lambda} \nabla_\mu\phi_2 J^\mu,
\end{align}
where \(J^\mu\) is some current, \(Q\) is some nonzero integer and both \(J^\mu\) and \(\Op\) contain heavy fields. In \Sec{subsec:chem-pot-pheno}, we show that the two options can be related through a field redefinition in the heavy fields. However, the required field redefinition may not be well-defined once the \(2\pi f_2\)-periodicity of \(\phi_2\) is considered, so these two options remain distinct.

In 4D EFT, the size of the chemical potential is constrained by the kinetic scale \(\sqrt{\dphicl}\approx 60H\). To see this, take the exponential interaction in \eq{eq:2-types} as an example. The chemical potential is given by
\begin{align}\label{eq:chem-pot-exp}
    \lambda &= \frac{Q\dot\phi_2}{f_2} \approx \frac{Q\dphicl}{f_2}.
\end{align}
From \eqs{eq:bi-axion-potential}{eq:2-types}, the EFT necessarily breaks down around the scale
\begin{align}\label{eq:axion-EFT-cutoff}
    \Lambda_a \simeq 4\pi \min\Bq{\frac{f_1}{N},\frac{f_2}{Q}}.
\end{align}
In general, we will consistently assume all Wilson coefficients for higher-dimensional operators to be weak below this scale: for an EFT operator \(\Op_{n,m}\), schematically \(\partial^m \phi^n\), this is
\begin{align}\label{eq:Wilson-unitarity-4D}
    c_{n,m} &\lesssim \frac{(4\pi)^{n-2}}{\Lambda_a^\Delta},&
    \Delta &= n_B +\frac{3}{2}n_F +m - 4,
\end{align}
where we include the \(4\pi\) counting rule. When applied to operators involving \(\nabla\phi_2\), we will consistently take the worst case where \(c_{n,m}\) is comparable to the bound \eq{eq:Wilson-unitarity-4D}. We then find that the 4D EFT breaks down if higher dimensional operators with more powers of
\begin{equation}\label{eq:higher-kinetic}
    16\pi^2 \frac{(\nabla\phi_2)^2}{\Lambda_a^4}  \supset 16\pi^2 \frac{\dphicl^2}{\Lambda_a^4} 
\end{equation}
become unsuppressed. To avoid this, the maximal chemical potential is constrained to be
\begin{equation}\label{eq:chem-max-naive}
    \lambda_{\max} \simeq \sqrt{4\pi\dphicl} \approx 200H.
\end{equation}
Compared to the qualitative estimates in \Cite{Creminelli:2003iq,Chen:2018xck}, here we included the \(4\pi\) power counting so that the chemical potential can be larger by approximately a factor of \(\sqrt{4\pi}\).
As we will see in \Sec{subsec:compact}, this na\"ive power counting will change in the context of higher-dimensional gauge fields where chemical potentials higher than this na\"ive upper-bound are possible.

\section{Inflation from 5D Gauge Fields}

\label{sec:A5-inflation}

Like other theories of axions, the theory of inflation from axions requires careful construction as any UV physics, including quantum gravity effects, can break the associated global symmetries and introduce a ``quality problem" of the delicate potential needed for the aligned axion mechanism, similar to the quality problem in QCD axions (see, e.g., \Cite{Hook:2018dlk} for review). In this paper, we are going to consider the higher-dimensional solution where the axions are identified as the fifth component of 5D gauge bosons:
\begin{align}
    \phi_k(x) &\sim \int \gf{A}_5^{(k)}(x,x_5) \dd{x_5},&
    k=1,2,
\end{align}
which is gauge invariant on a fifth dimensional circle or interval with Dirichlet boundary conditions for \(\gf{A}_\mu\).
The axion quality is then protected by gauge invariance and higher-dimensional locality. This idea of identifying PNGBs as the fifth component of gauge fields can be traced back to \Cite{Manton:1979kb,Hosotani:1983xw}, known as the \emph{Hosotani mechanism}. The idea of using a single higher-dimensional gauge field to model the inflaton is known as ``extranatural inflation" \cite{Feng:2003mk,Arkani-Hamed:2003xts}. However, the quantum gravity considerations from the weak gravity conjecture suggest multiple such axions paralleling the consideration in the previous section \cite{Arkani-Hamed:2006emk,delaFuente:2014aca}. Such higher-dimensional multi-axion models have also been considered in \cite{Bai:2014coa}.
In this work, we will mostly consider the case of a flat extra dimension for simplicity and defer the discussion of a warped extra dimension to \Sec{sec:discussion}.

\subsection{Gauge theory with an extra-dimensional interval}

\label{subsec:orbifolding}

While the original Hosotani mechanism and extranatural inflation focused on a circular fifth dimension, in this paper we will consider an interval, allowing the inflaton potential to be generated classically. Consider a 5D spacetime \(x^M = (x^\mu,x_5)\) with the fifth dimension compactified on a line segment of length \(L\), stabilized by the Goldberger-Wise mechanism \cite{Goldberger:1999uk}. We will work in the limit \(HL\ll 1\) where the spacetime metric is approximately given by
\begin{equation}
    \dd{s}^2_5 = \frac{-\dd{\eta}^2 +\dd{\vb{x}}^2}{\eta^2} +\dd{x}_5^2,
\end{equation}
whereas the exact geometry can be found in \cite{Kaloper:1998sw,Kaloper:1999sm}.

As a warm-up, we start with a single \(\mathrm{U}(1)\) gauge boson \(\gf{A}_M\) with the 5D action:
\begin{equation}\label{eq:single-gauge-action}
    S_{\gf{A}} = -\frac{1}{4} \int \sqrt{-g} \dd[4]{x} \int_0^L \dd{x_5} \gf{F}^{MN} \gf{F}_{MN}.
\end{equation}
To obtain a light scalar in the 4D spectrum, we choose the Dirichlet boundary conditions
\begin{equation}
    \eval{\gf{A}_\mu}_{x_5=0,L} = 0.
\end{equation}
The action of the gauge boson is now only invariant under 5D gauge transformation \(\gf{A}_M \mapsto \gf{A}_M +\partial_M \Lambda\) satisfying
\begin{equation}\label{eq:5D-gauge-constraint}
    \eval{\partial_\mu\Lambda}_{x_5=0,L} = 0.
\end{equation}
With these gauge transformations, one can gauge-fix \(\gf{A}_M\) such that \(\partial_5\gf{A}_5=0\). In this gauge, the 5D gauge field has the following Kaluza-Klein (KK) decomposition:
\begin{align}\label{eq:A5-KK}
    \gf{A}_\mu(x,x_5) &= \sqrt{\frac{2}{L}} \sum_{n=1}^\infty A^{(n)}_\mu(x) \sin\frac{n\pi}{L},&
    \gf{A}_5(x,x_5) &= \frac{1}{\sqrt{L}} \phi(x).
\end{align}
The spectrum of the 4D EFT thus contains a massless scalar \(\phi\) and vector bosons \(A_\mu^{(n)}\) of masses \(n\Mc\), where \(\Mc\coloneqq \pi/L\) defines the compactification scale.

In order to generate a potential for \(\phi=\sqrt{L} \gf{A}_5\), it is useful to further decompose a generic gauge transformation satisfying \eq{eq:5D-gauge-constraint} into a bulk gauge transformation plus two \(\mathrm{U}(1)\) transformations as
\begin{align}\label{eq:gauge-decomp}
    \Lambda(x,x_5) &= \lambda_0 \frac{L-x_5}{L} +\lambda_L \frac{x_5}{L} +\Lambda'(x,x_5),&
    \eval{\Lambda'}_{x_5=0,L}=0.
\end{align}
The bulk gauge transformation \(\Lambda'\) is eliminated by the gauge-fixing \(\partial_5\gf{A}_5=0\), while the two \(\mathrm{U}(1)\) survives as global symmetries, under which the scalar \(\phi\) transforms as
\begin{equation}
    \phi(x) \mapsto \phi(x) +\frac{\Lambda(x,L) -\Lambda(x,0)}{\sqrt{L}} = \phi(x) +\frac{\lambda_L -\lambda_0}{\sqrt{L}}.
\end{equation}
In order to obtain an inflaton potential, the \(\lambda_L-\lambda_0\) symmetry must be broken. Since 5D locality only allows breaking \(\lambda_0\) at \(x_5=0\) and \(\lambda_L\) at \(x_5=L\), the necessary condition to generate an inflaton potential on the interval is to break the \(\mathrm{U}(1)\) symmetry on both boundaries and connect these breaking by some ``messenger" bulk fields as illustrated below.

We will consider generating the inflaton potential at tree-level in a way analogous to the Goldberger-Wise mechanism. Consider adding a 5D charged scalar to \eq{eq:single-gauge-action}:
\begin{align}\label{eq:tree-A5-potential}
    S_{\gf{H}} &= -\int \sqrt{-g} \dd[4]{x} \int_0^L \dd{x_5} \p{\abs{D_M \gf{H}}^2 +m_H^2 \abs{\gf{H}}^2},
\end{align}
where \(\gf{H}\) has a unit charge under the \(\mathrm{U}(1)\) and the 5D gauge coupling is given by \(g_5\). To breaks the \(\mathrm{U}(1)\) symmetry on both boundaries, we assume that \(\gf{H}\) has inhomogeneous Dirichlet boundary conditions:
\begin{align}\label{eq:SB-bc}
    \eval{\gf{H}}_{x_5=0} &= v_{H0},& \eval{\gf{H}}_{x_5=L} &= v_{HL}.
\end{align}
In order to compute the effective potential, we consider constant \(\gf{A}_5\) and integrating out \(\gf{H}\) at tree level by solving its classical equation of motion. This results in a VEV of \(\gf{H}\):
\begin{align}\label{eq:H-VEV}
    \bk{\gf{H}(x_5)} &= v_{H0} \frac{\sinh m_H(L-x_5)}{\sinh m_HL} e^{ig_5\gf{A}_5x_5} +v_{HL} \frac{\sinh m_Hx_5}{\sinh m_HL} e^{-ig_5\gf{A}_5(L-x_5)}.
\end{align}
Plugging this into \eq{eq:tree-A5-potential}, we see that the equation of motion eliminates all bulk contribution and leaves only boundary terms:
\begin{align}\label{eq:H-potential}
    V_{\eff}(\gf{A}_5) 
    &= \frac{m_H}{\tanh m_HL} \p{\abs{v_{H0}}^2 +\abs{v_{HL}}^2} -\frac{m_H}{\sinh m_HL} \p{v_{H0} \bar v_{HL} e^{ig_5\gf{A}_5L} +\cc}.
\end{align}
Switching to the canonical 4D scalar \(\phi = \sqrt{L} \gf{A}_5\), we find that this generates  a single cosine potential for the scalar \(\phi\) as in \eq{eq:natural-inflation} with
\begin{align}\label{eq:potential-height}
    V_0 &= \frac{2m_H\abs{v_{H0} \bar v_{HL}}}{\sinh m_HL},\qquad
    f = \frac{1}{g_5\sqrt{L}}.
\end{align}
Similar models that generate the inflaton potential at tree-level can be found, e.g., in \cite{Deshpande:2019kjl,Petrossian-Byrne:2025mto,FernandezNavarro:2026bed}. Note that the potential height exhibit a Yukawa suppression \(e^{-m_HL}\) for \(m_H L\gg1\). This constraints the mass of \(\gf{H}\) to be less or comparable to \(\Mc\).

\begin{table}
    \centering
    \begin{tabular}{lccccc}\toprule
         Fields & \(\gf{A}_M^{(1)}\) & \(\gf{A}_M^{(2)}\) & \(\gf{H}_0\) & \(\gf{H}_1\) & \((X)\)\\\midrule
         \(\mathrm{U}(1)_1\) & 0 & 0 & 1 & \(-N\) & 0\\
         \(\mathrm{U}(1)_2\) & 0 & 0 & 0 & 1 & \(Q\)\\\bottomrule
    \end{tabular}
    \caption{The particle content of the higher-dimensional bi-axion model. The bi-axion potential \eq{eq:bi-axion-potential} is generated by \(\gf{H}_0\) and \(\gf{H}_1\) with \(\mathrm{U}(1)\)-breaking boundary conditions on both end points. Here we also include a heavy particle \(X\) charged under \(\mathrm{U}(1)_2\) as a target of the cosmological collider.}
    \label{tab:bi-axion-5D}
\end{table}

As mentioned before, one cannot model the inflaton with a single higher-dimensional gauge field without tensions with  the weak gravity conjecture, so a satisfactory model must contain at least two gauge fields, \(\gf{A}_M^{(1)}\) and \(\gf{A}_M^{(2)}\) with gauge couplings \(g_{51}\) and \(g_{52}\). The bi-axion potential in \eq{eq:bi-axion-potential} can then be achieved with two 5D scalars \(\gf{H}_1\) and \(\gf{H}_2\) with charge assignments shown in \tab{tab:bi-axion-5D}. Again, a trans-Planckian effective decay constant \(f_\eff\) can be obtained for \(f_1,f_2\ll\Mpl\) if \(N\) is large enough.

\subsection{Chemical potentials from minimal gauge interactions}

In the 5D gauge model of inflation, the slow-roll inflaton background naturally generates chemical potential for all particles charged under \(\gf{A}^{(2)}_M\). This is because the rolling inflaton background \(\dphicl\) corresponds to an  electric field in the fifth dimension:
\begin{equation}
    \dot\phi = \sqrt{L} \gf{F}^{(2)}_{05},
\end{equation}
resulting in different electrostatic energies for charged particles at different \(x_5\). When matching to the 4D EFT, a charge-\(Q\) KK mode \(\Psi_n\) with a 5D profile \(f_n(x_5)\) has a chemical potential given by
\begin{align}\label{eq:chem-pot-5D-gauge}
    \lambda_n &= \frac{Q\dphicl}{f_2} \frac{\bk{x_5}_n}{L},&
    \bk{x_5}_n \coloneqq \int_0^L x_5 \abs{f_n(x_5)}^2 \dd{x_5}.
\end{align}
Schematically, one can also regard these modes being localized around some \(\bk{x_5}\). The 5D gauge invariance together with the 5D locality then requires their 4D interactions to include by the Wilson line operators, e.g.,
\begin{equation}\label{eq:Wilson-line}
    \bar\Psi_m \Psi_n \exp\bq{iQ g_{52} \int_{x_5^{(n)}}^{x_5^{(m)}} \gf{A}_5^{(2)}(x,x_5) \dd{x_5}} = \bar\Psi_m \Psi_n \exp\bq{\frac{iQ\phi}{f_2} \frac{x_5^{(m)} - x_5^{(n)}}{L}},
\end{equation}
resembling the exponential interaction in \eq{eq:2-types} in the 4D EFT. Particle production through this mechanism has been studied on a circular fifth dimension in \Cite{Furuuchi:2015foh,Yamada:2024aca,Yamada:2025gvu}, but without producing an observable signals as we will do here in an extra-dimensional interval.

As an example, consider a charge-\(Q\) scalar \(X\) with brane-localized kinematic and mass terms at \(x_5=0\) with the following boundary conditions:
\begin{align}
    \eval{D_5X}_{x_5=0} &= 0,& \eval{X}_{x_5=L} &= v_{XL}.
\end{align}
The full Lagrangian is given by
\begin{align}\label{eq:brane-scalar-exp-model}
    S &= -\int \sqrt{-g} \dd[4]{x} \bq{ \int_0^L \dd{x_5} \p{\abs{D_M X}^2  +m_X^2 \abs{X}^2} +r \eval{\p{\abs{\nabla_\mu X}^2 +m^2_0 \abs{X}^2}}_{x_5=0} }.
\end{align}
We begin with a particularly simple and illustrative limit \(m_0,\lambda,r^{-1}\ll m_X,\Mc\) and leave the more general discussion to \Sec{eq:chem-KK}. In this limit, the 4D spectrum consists of a light mode approximately localized at \(x_5=0\) as well as heavy modes with approximate Dirichlet boundary condition at \(x_5=0\) \cite{Carena:2002me}. One can then separate this light mode, essentially \(X(0)\), from the heavy bulk degrees of freedom and integrate the latter out. The effective Lagrangian in 4D is similar to the one in \eq{eq:H-potential}:
\begin{align}\label{eq:brane-scalar-exp-Leff}
    \Ld_{\eff} &= -\abs{\nabla_\mu\chi}^2 -\p{m^2_0 +\frac{m_X}{r\tanh m_X L}} \abs{\chi}^2 -\frac{m_X \abs{v_{XL}}^2}{\tanh m_X L}\n
    &\peq  +\frac{m_X}{\sqrt{r} \sinh m_X L} \p{\bar v_{XL} \chi e^{iQ\phi_2/f_2} +\cc},
\end{align}
where \(\chi \coloneqq \sqrt{r} X(0)\) is the canonical scalar. Compared with \eq{eq:scalar-tree-model-exp-basis}, this reproduces the scalar model in the 4D EFT with the chemical potential given by \eq{eq:chem-pot-exp}, the physical mass given by
\begin{equation}
    m_\chi^2 = m^2_0 +\frac{m_X}{r\tanh m_X L}
\end{equation}
and the tadpole coefficient \(\alpha\) given by
\begin{equation}\label{eq:toy-tad-pole}
    \alpha = \frac{m_X \abs{v_{XL}}}{\sqrt{r} \sinh m_X L} = \frac{\abs{v_{XL}}}{\sqrt{r}} \begin{dcases}
        L^{-1}, & m_X \ll \Mc\\
        2m_X e^{-m_X L}, & m_X \gg \Mc.
    \end{dcases}
\end{equation}
To understand this model from our previous discussion, note that the interaction in \eq{eq:brane-scalar-exp-Leff} is exactly in the form \eq{eq:Wilson-line} where one of the \(\Psi_n\) is simply \(\chi\) and the other one is replaced by the VEV \(v_{XL}\), which acts as a reservoir of charge that allow single charged particles to be created and annihilated. The bulk part of \(X\) then acts as a mediator that allows \(\chi\) at \(x_5=0\) to leak into the bulk and overlap with the reservoir at \(x_5=L\).

\subsection{Chemical potentials from non-minimal interactions}

\label{subsec:nonmin}

Besides minimal gauge interactions, the chemical-potential interaction \eq{eq:4D-chem-potential-2} can also arise from non-minimal coupling between the 5D gauge field and matter. At linear order in \(\gf{F}^{(2)}_{MN}\), these include
\begin{itemize}
    \item bulk interactions \(\dfrac{\order(4\pi)}{\LambdafD^\Delta} \gf{F}^{(2)}_{MN} \Op^{MN}_{\Delta+5/2}\), and
    \item brane-localized interactions \(\dfrac{\order(4\pi)}{\LambdafD^\Delta} \gf{F}^{(2)}_{\mu5} \Op^\mu_{\Delta+3/2}\),
\end{itemize}
where \(\LambdafD\) is the cutoff scale of the 5D EFT and \(\Op^{MN}_n\) and \(\Op^\mu_n\) are 5D and 4D operators of dimension \(n\) that are bilinear in matter fields, respectively.\footnote{When massive vector fields are involved, the power counting should be performed using the St\"uckelberg replacement \(A_\mu \mapsto \frac{1}{m_A} \partial_\mu \pi\) if the longitudinal mode is involved. For example, the operator \(\chi B^\mu\) from a brane-localized scalar \(\chi\) and massive vector \(B^\mu\) should be regarded as dimension-3 rather than dimension-2.}
Compared to minimal gauge interactions, such non-minimal interactions can exist for particles neutral under the inflationary gauge field \(\gf{A}^{(2)}_M\). 

The size of the chemical potential is given by
\begin{align}\label{eq:chem-pot-non-minimal}
    \lambda &= \order(4\pi)\cdot \frac{\dphicl \Mc^{\Delta-1}}{\sqrt{\pi} \LambdafD^\Delta},\qquad
    \Delta\ge\frac{3}{2}.
\end{align}
Compared with \(\lambda=\dphicl/\Lambda\) in 4D EFT, chemical potentials from non-minimal interactions are further suppressed by powers of \(\Mc/\LambdafD\) after identifying \(\Lambda=\LambdafD\) and are thus expected to be smaller than the na\"ive estimation \eq{eq:chem-max-naive}. Nonetheless, they are worth discussing as they draw clear connections to existing 4D chemical potential models in the literature discussed in \Sec{subsec:chem-pot-pheno}. Here we we provide two example models and leave further discussions in \Sec{sec:chem-nonmin-p}:

\paragraph{U(1) chemical potential for brane scalar.}
This model consists of a complex scalar \(\chi\) localized on the \(x_5=0\) boundary, non-minimally coupled to the 5D gauge field through a brane-localized operator as follow:
\begin{equation}\label{eq:brane-scalar-model}
    S = -\int_{x_5=0} \Bq{ \abs{\nabla_\mu\chi}^2 +m^2 \abs{\chi}^2 +\frac{c}{\LambdafD^{3/2}} \gf{F}^{(2)}_{\mu 5} J_{(\chi)}^\mu +\frac{c'}{\LambdafD^3} \bq{\gf{F}^{(2)}_{\mu 5}}^2 \abs{\chi}^2 +\alpha \chi +\bar\alpha \bar\chi} \dd[4]{x}.
\end{equation}
After substituting in \eq{eq:A5-KK}, this becomes
\begin{equation}\label{eq:brane-scalar-4D}
    \Ld_{\mathrm{4D}} = -\abs{\nabla_\mu\chi}^2 -m^2 \abs{\chi}^2 -\frac{c}{\LambdafD^{3/2}L} \partial_\mu\phi_2 J_{(\chi)}^\mu -\frac{c'}{\LambdafD^2 L^2} (\nabla_\mu\phi_2)^2 \abs{\chi}^2 -\alpha \chi -\bar\alpha \bar\chi,
\end{equation}
which reproduces the scalar model \eq{eq:scalar-tree-model-chem-basis} when \(c'=c^2\) with \(\Lambda=\Lambdaeff\) given by
\begin{equation}\label{eq:Lambdaeff}
    \Lambdaeff \coloneqq c^{-1} \LambdafD^{3/2}L^{1/2}.
\end{equation}
The generic case \(c'\ne c^2\) is also discussed in \cite{Bodas:2020yho} and the result is qualitatively the same.

\paragraph{Helicity-dependent chemical potential for bulk vector boson.}
This model consists of another bulk gauge boson \(\gf{B}_M\) with Neumann boundary conditions coupled to \(\gf{A}^{(2)}_M\) through a mix Chern-Simons (CS) interaction in 5D:
\begin{align}\label{eq:KK-gauge-model}
    S_1 &= \int \bq{ -\frac{1}{4}\gf{G}^{MN} \gf{G}_{MN} +\frac{1}{8\Lambda_{\text{CS}}^{3/2}} \epsilon^{MNPQR} \gf{A}_M^{(2)} \gf{G}_{NP} \gf{G}_{QR} } \dd[5]{x},& \frac{1}{\Lambda_{\text{CS}}^{3/2}} &\coloneqq \frac{g_{52} g_5'^2 k_{\text{CS}}}{4\pi^2},
\end{align}
where \(\gf{G}_{MN}\) and \(g_5'\) are the field strength and the gauge coupling of \(\gf{B}_M\) respectively, and \(k_{\text{CS}}\) is the integer CS level. To have maximal chemical potential, we will assume \(k_{\text{CS}}\) is such that \(\Lambda_{\text{CS}} \approx \LambdafD\). Consider the target to be the lowest KK mode of \(\gf{B}_M\), namely \(B_\mu\). In order to generate a mass for \(B_\mu\), we also include a Higgs field \(\Sigma\) living in the bulk:
\begin{align}
    S_2 &= -\int \Bq{ \abs{D_M\Sigma}^2 -m_\sigma^2 \abs{\Sigma}^2 -\lambda_5 \abs{\Sigma}^4 +\frac{c}{\LambdafD^3} \abs{\Sigma}^2 \bq{\gf{F}_{MN}^{(2)}}^2 },
\end{align}
where \(\Sigma\) has Neumann boundary conditions and we include the leading-order interactions between \(\Sigma\) and \(\gf{A}_{M}\) as suggested in \cite{Wang:2020ioa}. The effective 4D Lagrangian is now given by
\begin{align}\label{eq:KK-gauge-4D}
    \Ld_{\mathrm{4D}} &\supset -\frac{1}{4} G^{\mu\nu} G_{\mu\nu} -\frac{1}{8\Lambdaeff} \epsilon^{\mu\nu\rho\sigma} \phi_2 G_{\mu\nu} G_{\rho\sigma}\n
    &\peq -\abs{D_\mu \sigma}^2 +\p{m^2_\sigma -\frac{c}{\Lambdaeff'^2} \abs{\nabla_\mu\phi_2}^2} \abs{\sigma}^2 -\lambda \abs{\sigma}^4,
\end{align}
the same as the Lagrangian discussed in \cite{Wang:2020ioa} with \(\Lambdaeff\coloneqq \Lambda_{\text{CS}}^{3/2}L^{1/2}\), \(\Lambdaeff' \coloneqq \LambdafD^{3/2}L^{1/2}\) and the scalar quartic coupling \(\lambda \coloneqq \lambda_5/L\). The leading-order cosmological collider signal clearly originates from one-loop productions of the heavy vector bosons.

\subsection{Constraints on 5D EFT of inflation}

\label{subsec:constraints}

In previous discussions, we found that the size of chemical potentials is determined by three different scales: the compactification scale \(\Mc\), the decay constant \(f\) emergent from 5D gauge theory and the cutoff scale \(\LambdafD\) of the 5D EFT. Meanwhile, among various types of targets we can consider, we are interested in KK excitations with masses comparable to \(\Mc\). In this section, we will derive the constraints on the higher-dimensional gauge theory of inflation. These constraints will be used to find the maximal chemical potentials in the two different scenarios as well as whether KK excitations can be produced.

\subsubsection{Slow-roll inflation}

We first begin with basic requirements from slow-roll inflation. In the bi-axion model, successful slow-roll inflation at the highest Hubble scale \(H\approx\qty{6e13}{GeV}\) requires
\begin{equation}\label{eq:f_eff-2}
    f_{\eff} \simeq Nf_2 \approx 10\Mpl 
\end{equation}
and
\begin{align}\label{eq:inflaton-potential}
    V_0 &\gtrsim \frac{3}{2} \Mpl^2H^2 \approx (220H)^4,&
    V_1 &\gtrsim \frac{1}{N} V_0,
\end{align}
where the second condition is needed to ensure a winding trajectory near \eq{eq:bi-axion-traj} exists over the whole \(\Delta\phi_1 = \pi f_1\).

\subsubsection{EFT constraints}

Since 5D gauge theory is non-renormalizable, the higher-dimensional theory of inflation is only valid up to a cutoff scale \(\LambdafD\), which we will specify as the scale where new physics must arise. Below this scale, we expect the EFT to be weakly-interacting: for an EFT operator \(\Op_{n,m}\), schematically \(\partial^m \Phi^n\), its Wilson coefficient should be smaller than\footnote{In 5D, the angular loop factor is slightly different from the familiar \(16\pi^2\) in 4D, though the difference is not going to be important for our discussion.}
\begin{align}\label{eq:Wilson-unitarity}
    c_{n,m} &\lesssim \frac{(4\pi)^{n-2}}{\LambdafD^\Delta},&
    \Delta &= \frac{3}{2} n_B +2n_F +m - 5.
\end{align}
In particular, applying this to the gauge interactions in the bi-axion model \tab{tab:bi-axion-5D} with the addition of a target heavy particle of charge \(Q\) under \(\mathcal{A}_M^{(2)}\) gives
\begin{align}\label{eq:5D-gauge}
    \LambdafD &\lesssim \min\Bq{\Lambda_{g1},\Lambda_{g2}},&
    \begin{dcases}
        \Lambda_{g1} = \frac{16\pi^2}{g^2_{51} N^2},\\
        \Lambda_{g2} = \frac{16\pi^2}{g^2_{52} Q^2}.
    \end{dcases}
\end{align}
This can be rewritten using the decay constants \(f_{1,2}\) as
\begin{equation}\label{eq:5D-gauge-2}
    \LambdafD \lesssim \frac{16\pi^3}{\Mc} \cdot \min\Bq{ \frac{f^2_1}{N^2}, \frac{f^2_2}{Q^2} }.
\end{equation}
In the limit \(\Mc\sim\LambdafD\), we see that this reproduces the na\"ive estimation \eq{eq:axion-EFT-cutoff} based on the 4D EFT with \(\LambdafD \simeq \Lambda_a \) up to \(\order(1)\) factors.

Besides an upper-bound on \(\LambdafD\), an EFT argument can also be used to constrain the maximal VEV of the 5D scalar \(\gf{H}\) and thus the maximal potential height in \eqs{eq:potential-height}{eq:bi-axion-potential}: since for any EFT operator \(\Op_0\), there is an associated series of higher-dimensional operators \(\Op_n = \Op_0 \abs{\mathcal{H}}^{2n}\), EFT control requires \(c_n \abs{\bk{\gf{H}}}^{2n}\) decreasing over \(n\).
In the most conservative scenario, one assumes that all Wilson coefficients saturates the unitarity bound \eq{eq:Wilson-unitarity}, leading to
\begin{align}\label{eq:potential-bound-s}
    \bk{\gf{H}_{0,1}} &\lesssim \frac{1}{4\pi}\LambdafD^{3/2}, &
    V_{0,1} \lesssim \frac{2}{\pi} \Mc \bk{\gf{H}_{0,1}}^2 \lesssim \frac{1}{8\pi^3} \Mc \LambdafD^3.
\end{align}
This turns out to be too stringent as \(\gf{H}_0\), the charged particle responsible for the inflaton potential, is \(1/N\) less charged than \(\gf{H}_1\). It is then consistent to assume that an EFT operator with \(n_0\) powers of \(\gf{H}_0\) has a Wilson coefficient \(1/N^{n_0/2}\) less than the na\"ive expectation \eq{eq:Wilson-unitarity}, leading to
\begin{align}\label{eq:potential-bound-w}
    \bk{\gf{H}_0} &\lesssim \frac{\sqrt{N}}{4\pi}\LambdafD^{3/2}, &
    V_0 &\lesssim \frac{2}{\pi} \Mc \bk{\gf{H}_0}^2 \lesssim \frac{N}{8\pi^3} \Mc \LambdafD^3.
\end{align}

While perturbative unitarity constrains the theory from being strongly-interacting, there are also constraints on how \emph{weak} the gauge coupling \(g_5\) can be from quantum gravity considerations, namely the \emph{Weak Gravity Conjecture} (WGC) \cite{Arkani-Hamed:2006emk}. The conjecture states that there always exists ``elementary" charged objects with mass-to-charge ratio smaller than the corresponding extremal black holes in order for them to decay.
Applying this to the magnetic dual of the two 5D \(\mathrm{U}(1)\) gauge theories implies \cite{delaFuente:2014aca}
\begin{align}
    \sqrt{\frac{\LambdafD^2}{g_{51}^2} +\frac{\LambdafD^2}{g_{52}^2}} &\lesssim \frac{\Mpl}{\sqrt{L}},
\end{align}
or in terms of decay constants \(f_{1,2}\):
\begin{equation}\label{eq:WGC-2}
    \sqrt{f_1^2 +f_2^2} \lesssim \frac{1}{\pi} \p{\frac{\Mc}{\LambdafD}} \Mpl.
\end{equation}
This reduces back to \(f_{1,2}\lesssim\Mpl\) when \(\Mc\simeq \LambdafD\), while a hierarchy between \(\LambdafD\) and \(\Mc\) leads to a tighter upper-bound on \(f_{1,2}\).

It is worth mentioning that ref. \cite{Saraswat:2016eaz} has discussed a loophole of this argument. The counter-example model also involves two \(\mathrm{U}(1)\) gauge fields but with a \emph{bulk} Higgs of quantum numbers similar to the \(\gf{H}_1\) in \tab{tab:bi-axion-5D}. If we place this model also in 5D and add the same \(\gf{H}_0\) to generate a potential for the inflaton, in the limit where the bulk Higgs \(\gf{H}_1\) is heavy, the bi-axion potential is instead approximately given by
\begin{align}
    V(\phi) &\simeq V_0 \p{1 -\cos\frac{\phi_1}{f_1}} +\frac{\abs{\bk{\gf{H}_1}}^2}{L} \p{-\frac{N\phi_1}{f_1} +\frac{\phi_2}{f_2} +2\pi k}^2,& k\in\Z.
\end{align}
The second term comes from the mass term for \(-Ng_{51}\gf{A}_1 +g_{52}\gf{A}_2\) but with a monodromy structure as required by the periodicity of the axions. Yet for \(N\gg1\), integrating out the heavy mode projects out the same trajectory \eq{eq:bi-axion-traj} and leads to similar phenomenology in 4D. The only difference is that \(\gf{H}_1\) (or the physical Higgs \(h_1\)) is no longer required to have a mass comparable to \(\Mc\) to avoid the Yukawa suppression in \eq{eq:potential-height}, which can be a desirable feature as \(\gf{H}_1\) is more natural to have a mass around \(\LambdafD\) when \eq{eq:5D-gauge-2} is around saturation. For simplicity, we will not pursue this option here.

\subsubsection{Quality problem}

Finally, solving the quality problem of the inflaton potential requires a minimal hierarchy between \(\LambdafD\) and \(\Mc\). To see this, assuming some new particle with charge \((n_1, n_2)\) arises at the cutoff scale \(\LambdafD\) with the same symmetry-breaking boundary conditions as \eq{eq:SB-bc}, it will then generate a correction to the inflaton potential as follow
\begin{align}\label{eq:inflaton-potential-correction}
    \delta V &\simeq 4\LambdafD \abs{v_\Lambda}^2 e^{-\LambdafD L} \cos\p{\frac{n_1\phi_1}{f_1} +\frac{n_2\phi_2}{f_2} +\theta}\n
    &\sim \delta V_0 \cos\bq{ \frac{\p{Nn_2 +n_1} \phi_L}{f_\eff} +\theta},&
    \delta V_0 &\simeq 4\LambdafD \abs{v_\Lambda}^2 e^{-\pi\LambdafD/\Mc}.
\end{align}
In general, we expect that \(Nn_2 +n_1=\order(N)\gg1 \), so this contributes a periodic modulation on top of the slow-roll potential. The correction from this to the power spectrum has been estimated in \Cite{Flauger:2009ab,Choi:2015aem} and compared to CMB data prior to Planck 2018, and a more recent fit can be found in \Cite{Calderon:2025xod}. According to these estimates, \eq{eq:inflaton-potential-correction} generates a correction to the power spectrum in the following form
\begin{align}
    \frac{\delta{\Ps}(k)}{\Ps(k)} &= \delta n_s \cos\p{\omega \log\frac{k}{k_*} +\varphi_*},&
    \left\{\begin{aligned}
        \delta n_s &= \sqrt{2\pi\omega} \frac{\delta V_0}{\dphicl^2} = \frac{3\sqrt{2\pi\omega}}{2\epsilon_H} \frac{\delta V_0}{V_0},\\
        \omega &= \frac{(Nn_2 +n_1)\dphicl}{Hf_\eff}.
    \end{aligned}\right.
\end{align}
For \(1<\omega<100\), the constraint on \(\delta n_s\) from Planck 2018 is given by \(\delta n_s<0.021\) \cite{Calderon:2025xod}, so
\begin{align}\label{eq:mod-constraint}
    \frac{\delta V_0}{V_0} \simeq \frac{4\LambdafD \abs{v_\Lambda}^2 e^{-\pi\LambdafD/\Mc}}{\frac{3}{2}\Mpl^2 H^2} \lesssim \frac{\num{2.2e-5}}{\sqrt{\omega}}.
\end{align}
For slow-roll potential saturating \eq{eq:potential-bound-w} and \(\abs{v_\Lambda}\) as large as \(\bk{\gf{H}_0}\), this suggests 
\begin{equation}\label{eq:quality-bound}
    \LambdafD \gtrsim \p{4.5 +0.4 \log_{10} \omega}\Mc.
\end{equation}
Since the \(f_\eff\) constraint \eq{eq:f_eff-2} and the WGC constraint \eq{eq:WGC-2} together imply
\begin{align}\label{eq:L/M-maximal}
    \Mc &\gtrsim \p{\frac{10\pi}{N}} \LambdafD,
\end{align}
solving the quality problem consistently with the WGC constraint requires an uncomfortably large \(\mathrm{U}(1)_2\) charge \(N\gtrsim 140-170\). As we will see in the next sections, smaller charges are possible in extended models with more axions.

\section{Probing the Extra Dimension with Chemical Potentials}

Given the higher-dimensional model of inflation, we then ask whether cosmological collider signals from KK excitations can be produced by the chemical potential mechanism. Since KK excitations have masses of \(\order\p{\Mc}\), this requires a chemical potential \(\lambda\) higher than the compactification scale.
In this section, we will first show that this is possible for chemical potentials from minimal gauge interactions and then explore the phenomenology of such models.

\subsection{Observable compactification scale and multi-axion extension}

\label{subsec:compact}

With all the constraints in \Sec{subsec:constraints}, we can now check whether \(\lambda\gtrsim\Mc\) is parametrically possible. In the simplest case with \(f_1=f_2\), it happens that both the minimal \(\Mc\) and the maximal \(\lambda\) are achieved simultaneously by saturating the \(f_\eff\) constraint \eq{eq:f_eff-2}, the EFT constraints eqs. \eqref{eq:5D-gauge-2}, \eqref{eq:potential-bound-w} and the WGC constraint \eq{eq:WGC-2}. The corresponding values for all the parameters are summarized in \tab{tab:Mc-min} with \# axion\:=\:2, where we found that the ratio \(\lambda/\Mc\) can be as large as \(9\). 
Note that the maximal chemical potential can actually exceed the na\"ive 4D bound in \eq{eq:chem-max-naive} because in the 5D EFT, the potentially dangerous higher-order operators involving more powers of \(\dphicl\) as in \eq{eq:higher-kinetic} are instead controlled by
\begin{align}
    16\pi^2 \frac{\gf{F}_{MN}^2}{\LambdafD^5} &\supset 16\pi \frac{\Mc\dphicl^2}{\LambdafD^5} \lesssim 16\pi \p{\frac{\Mc}{\LambdafD}}^3 \frac{\dphicl^2}{\Lambda_a^4},
\end{align}
with \(\Lambda_a\) defined in \eq{eq:axion-EFT-cutoff}. With the extra suppression from \((\Mc/\LambdafD)^3\), it is straightforward to verify that this term is always less than \(\num{e-2}\).

\begin{table}
    \centering
    \begin{tabular}{cccccc}\toprule
         \# axions & \(N\) & \(\Mc/H\) & \(\LambdafD/H\) & \(f/H\) & \(\lambda/H\)\\\midrule
         2 & 249 & 61 & 340 & 1610 & 560\\
         3 & 39 & 29.3 & 810 & 267 & 520\\
         4 & 15.3 & 21.3 & 1220 & 111 & 500\\\bottomrule
    \end{tabular}
    \caption{Parameters found by saturating eqs. \eqref{eq:f_eff-2}, \eqref{eq:5D-gauge-2}, \eqref{eq:WGC-2} and \eq{eq:potential-bound-w}, corresponding to the minimal compactification scale \(\Mc\),  the maximal chemical potential \(\lambda\) and the maximal value for the charge \(N\). For simplicity, we have considered \(f_1=\cdots=f_n=f\) and \(N_1=\cdots=N_{n-1}=N\).}
    \label{tab:Mc-min}
\end{table}

While such a large value of \(\lambda/\Mc\) is promising, the bi-axion model only marginally satisfies all the constraints as the marginalization suggests that \(\LambdafD/\Mc\le5.6\) according to \tab{tab:Mc-min}, whereas the remaining bound \eq{eq:quality-bound} from solving the quality problem requires \(\LambdafD/\Mc \gtrsim 4.5-5.3\).
Even this marginal success is doubtful in the face of the various \(\order(1)\) uncertainties in the EFT constraints eqs. \eqref{eq:5D-gauge-2}, \eqref{eq:potential-bound-w} and the WGC constraint \eq{eq:WGC-2}.
Furthermore, solving the quality problem with the WGC constraint already requires \(\mathcal{H}_1\) to have a very large \(\mathrm{U}(1)_2\) charge \(N\gtrsim\order(100)\), which seems bizarre.
Luckily, both concerns can be alleviated by extending this model with more axions and the following scalar potential \cite{Choi:2014rja}:
\begin{align}
    V(\phi) &= V_0 \p{1 -\cos\frac{\phi_1}{f_1}} +V_1 \bq{1 -\cos\p{ -\frac{N_1\phi_1}{f_1} +\frac{\phi_2}{f_2}}}\n
    &\peq +\cdots +V_n \bq{1 -\cos\p{-\frac{N_{n-1} \phi_{n-1}}{f_{n-1}} +\frac{\phi_n}{f_n}}}.
\end{align}
The higher-dimensional realization is straightforward. Integrating out the heavy modes then yields the effective potential for the lightest mode \(\phi_L \simeq \phi_n\):
\begin{align}\label{eq:feff-multi-axion}
    V(\phi_L) &= V_0 \p{1 -\cos\frac{\phi_L}{f_{\eff}}},&
    f_{\eff} \simeq N_1N_2\cdots N_{n-1} f_n.
\end{align}
Since the WGC constraint now only implies \(N_\eff = N_1 N_2\cdots N_{n-1} \gtrsim \order(100)\), achievable with much smaller individual charges, the constraint \eq{eq:5D-gauge-2} from perturbative unitarity is also loosen. We computed the marginalized parameters of the same set of constraints for tri-axion and quad-axion models with \(f_1=\cdots=f_n=f\) and \(N_1=\cdots=N_{n-1}=N\) for simplicity. The results are also included in \tab{tab:Mc-min}. It is clear that the marginalized parameters now allow larger hierarchy between \(\LambdafD\) and \(\Mc\) expected by the quality arguments while the required charges \(N\) are lowered down to \(\order(10)\).

Here we provides a benchmarks consistent with all the constraints in \Sec{subsec:constraints} based on the tri-axion model:
\begin{align}\label{eq:KK-benchmark}
    N &= 27,&
    \Mc &= 120H,&
    \LambdafD &= 750H,&
    f &= 560H,&
    \lambda &\le 174H.
\end{align}
It is straightforward to verify that this benchmark satisfies the EFT constraints eqs. \eqref{eq:5D-gauge-2}, \eqref{eq:potential-bound-w} and the WGC constraint \eq{eq:WGC-2} all by more than a factor of \(2\).

\subsection{Particle production from Schwinger effects}

\label{eq:chem-KK}

When \(\lambda\gtrsim\Mc\) for a charged particle, it turns out that the 5D electric field \(\gf{F}_{05}\) also has a strong impact on the 5D profiles of the charged field.
To see this, consider the same model in \eq{eq:brane-scalar-exp-model} but with \(r\) negligible and arbitrary boundary conditions.
Under the slow-roll background, the zeroth order equation of motion is given by
\begin{align}
    \bq{\eta^2\partial_\eta^2 -2\eta \partial_\eta -\p{\partial_5 -\frac{i\lambda_0 t}{L}}^2 +k^2\eta^2 +m_X^2} X_{\vb{k}} &= 0,& \lambda_0 &\coloneqq Qg_5\sqrt{L}\dphicl= \frac{Q\dphicl}{f_2}.
\end{align}
With the adiabatic approximation, the general solution can be expanded as
\begin{equation}\label{eq:KK-decomposition-E}
    X_{\vb{k}}(\eta,x_5) \simeq \sum_{n=1}^\infty \frac{\chi_{n\vb{k}}}{\sqrt{2\omega_n(-k\eta)}} (-\eta)^{3/2-i\lambda_0 \p{1 -x_5/L}} e^{-i\int \omega_n(-k\eta) \dd{\log(-\eta)}} f_n(x_5;-k\eta),
\end{equation}
where the profiles \(f_n(x_5;p)\) are solutions to
\begin{align}\label{eq:dphi-mode-function}
    \bq{-\partial_5^2 -\p{\omega_n +\lambda_0 \frac{L-x_5}{L}}^2 +p^2 +\mu_X^2} f_n(x_5;p) &= 0,&
    \mu_X \coloneqq \sqrt{m_X^2 -\frac{9}{4}}.
\end{align}
The solutions can be expressed in terms of parabolic cylinder functions. Note that in \eq{eq:KK-decomposition-E}, the frequencies are \(x_5\)-dependent and we have defined \(\omega_n\) to be the frequency at \(x_5=L\).

\begin{figure}
    \centering
    \includegraphics[scale=0.78]{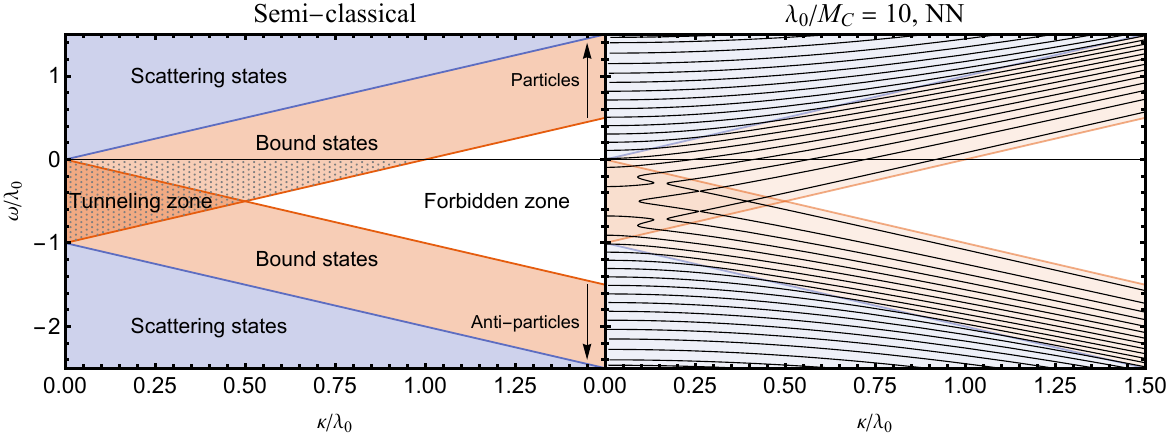}
    \caption{(\emph{Left}) the eigenvalues \((\omega,\kappa)\) in the semi-classical limit. During inflation, an eigenmodes starts at \(\kappa\to\infty\) and evolves down to \(\kappa=m\) at late-time. The existence of the electric field leads to negative energies for particle solutions (shaded region) as well as eigenvalue crossing (tunneling zone) at late-time if \(2\mu_X < \lambda_0\). (\emph{Right}) an explicit example of the spectrum from numerical method for \(\lambda_0/\Mc=10\) with Neumann boundary conditions on both ends.}
    \label{fig:KK-Schwinger}
\end{figure}

\Eq{eq:dphi-mode-function} is analogous to a time-independent Schr\"odinger equation
\begin{align}\label{eq:SE}
    \mathscr{H} \psi_n &= E_n \psi_n,& \left\{\begin{aligned}
        \mathscr{H} &= -\partial_5^2 -\p{\omega +\lambda_0 \frac{L-x_5}{L}}^2,\\
        E_n &= -\p{p_n^2 +\mu_X^2} \eqqcolon -\kappa_n^2,
    \end{aligned}\right.
\end{align}
where we define \(\kappa\coloneqq\sqrt{p^2 +\mu_X^2}\) for later convenience. It is easier to first understand the phenomenology in the semi-classical limit \(\lambda_0L,\omega L\gg1\). In this limit, the spectrum is almost continuous and we can divide the \((\omega,\kappa)\)-space into 4 regions:
\begin{enumerate}
    \item \emph{Forbidden zone}: \(-\kappa<\omega<\kappa-\lambda_0\) from \(E < \min V(x_5)\). This region is forbidden both classically and quantum mechanically for usual boundary conditions. At early time during inflation where \(p = -k\eta\to\infty\), the forbidden zone separates positive frequency modes from negative frequency modes, allowing one to clearly define the Bunch-Davies vacuum as the initial condition: \(f_n(x_5)\) with \(\omega_n>0\) at \(\kappa\to\infty\) are identified as particle solutions while those with \(\omega_n<0\) at \(\kappa\to\infty\) are identified as anti-particle solutions. 
    
    \item \emph{Scattering states}: \(\omega>\kappa\) or \(\omega<\kappa-\lambda_0\) from \(E > \max V(x_5)\). In this case, the (anti-)particles can propagate classically through all \(0<x_5<L\).
    
    \item \emph{Bound states}: \(\kappa-\lambda_0<\omega<\kappa\) or \(-\kappa-\lambda_0<\omega<-\kappa\) from \(\min V(x_5)<E<\max V(x_5)\). In this case, we have either particle solutions localized near \(x_5=0\) for \(\kappa-\lambda_0<\omega<\kappa\) [\(E>V(0)\)], or anti-particle solutions near \(x_5=L\) for \(-\kappa-\lambda_0<\omega<-\kappa\) [\(E>V(L)\)].
    
    \item \emph{Tunneling zone}: \(\kappa-\lambda_0<\omega<-\kappa\). In this region, the na\"ive particle spectrum and anti-particle spectrum intersect at some values of \((\omega_*,\kappa_*)\). These would-be eigenvalue crossing are actually avoided due to quantum tunneling. To see this, note that the equation near the crossing point is schematically described by the two-level system
    \begin{align}\label{eq:avoided-crossing}
        \begin{bmatrix}
         -\kappa_*^2-c_* (\omega-\omega_*) & \Delta\\
         \bar\Delta & -\kappa_*^2 +c_* (\omega -\omega_*)
        \end{bmatrix} \Psi &= -\kappa^2 \Psi,
    \end{align}
    where \(\Delta\) can be estimated through the WKB approximation:
    \begin{equation}\label{eq:pair-rate}
        \log \Delta \simeq -\int_{x_{5,\min}}^{x_{5,\max}} \sqrt{\kappa^2 -\p{\omega +\lambda_0 \frac{L-x_5}{L}}^2} \dd{x_5} = -\frac{\pi\kappa^2L}{2\lambda_0}.
    \end{equation}
    The dispersion relation near the crossing is thus given by
    \begin{equation}
        \p{\kappa^2 -\kappa_*^2}^2 -c_*^2 (\omega -\omega_*)^2 = \abs{\Delta}^2.
    \end{equation}
    In particular, for \(\abs{\kappa^2 -\kappa_*^2}<\abs{\Delta}\), the mode function grows exponentially as \(e^{\gamma t/2}\) with \(\gamma \propto \abs{\Delta}\). Physically, such instability is induced by the classic Schwinger pair production from the electric field along the fifth dimension. This can be seen explicitly from the production rate of the charged particle:
    \begin{equation}
        \Gamma \propto \gamma^2 \propto \exp\bq{-\frac{\pi L\p{\vb{p}^2 +\mu_X^2}}{\lambda_0}} = \exp\bq{-\frac{\pi\p{\vb{p}^2 +\mu_X^2}}{Qg_5 \abs{\gf{F}_{05}}}},
    \end{equation}
    which contains the same exponential factor as in the usual Schwinger mechanism \cite{Sauter:1931zz,Heisenberg:1936nmg,Schwinger:1951nm}. This effect remains active for fixed electric field \(\gf{F}_{05}\) in the limit \(L\to\infty\).
\end{enumerate}

Finally, if explicit \(\mathrm{U}(1)\) breaking happens at either 4D boundaries, charged particles can also be created or annihilated there. Assuming the breaking happens at \(x_5=L\), a charged particle \(X\) can be produced on-shell simultaneously with an inflaton if \(\omega_n(p) +p = 0\) has a solution for \(p\) and \(\omega_n\) from particles solutions. This is possible if \(\kappa-\lambda_0<\omega<0\). Note that since this process can only happen for particle solutions bounded near \(x_5=0\), the associated amplitude is also suppressed by a tunneling rate similar to the one in \eq{eq:pair-rate}. For example, if such breaking is induced by a non-zero VEV of \(X\) at \(x_5=L\), the creation amplitude will be roughly proportional to the overlap between the 5D profile of the bound state and the VEV. The latter is also obtained from \eq{eq:dphi-mode-function} but with \(\omega = p = 0\). It turns out that such overlap is dominated by the tunneling of the VEV from \(x_5=L\) to \(x_5=0\), for which the WKB approximation gives
\begin{equation}\label{eq:tree-exp-sup}
    \exp\bq{ -\int_{x_{5,\min}}^L \sqrt{\mu_X^2 -\lambda_0^2 \p{\frac{L-x_5}{L}}^2} \dd{x_5}} = \exp\p{-\frac{\pi\mu_X^2L}{4\lambda_0}}.
\end{equation}
The discussions above are summarized in \fig{fig:KK-Schwinger}. In the semi-classical limit \(L\to\infty\), the conditions for particle productions are
\begin{align}
    \lambda_0 &> \mu_X, & \text{for boundary production, and}\\
    \lambda_0 &> 2\mu_X, & \text{for pair production}.
\end{align}

\begin{figure}
    \centering
    \begin{tabular}{ll}
        \includegraphics[scale=0.8]{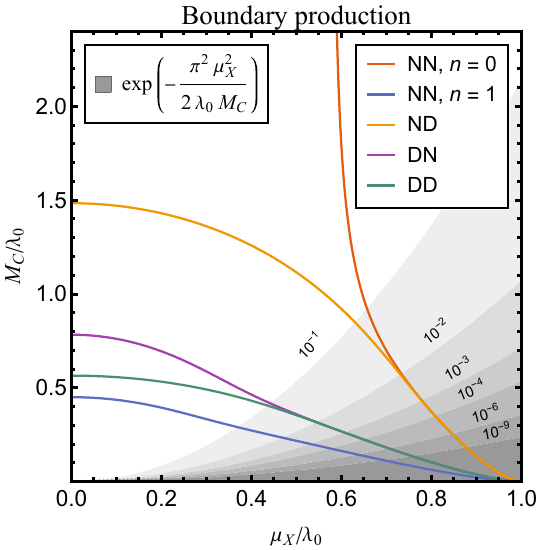} &
        \includegraphics[scale=0.8]{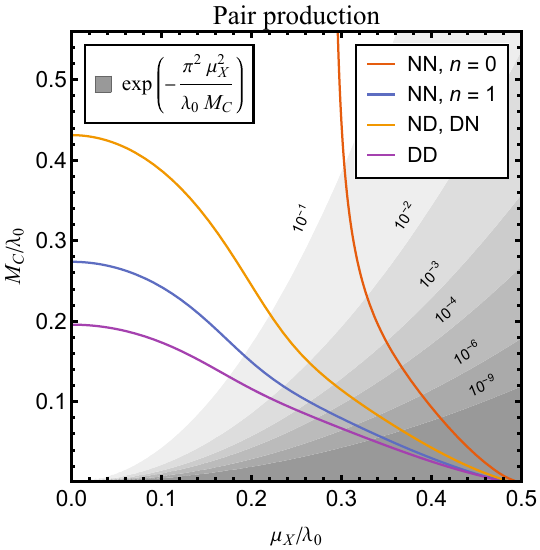}
    \end{tabular}
    \caption{The parameter regions for the two production mechanisms under different boundary conditions. The boundary conditions are labeled as ``D" for Dirichlet and ``N" for Neumann with the first letter for \(x_5=0\) and the second one for \(x_5=L\). For the ``NN" boundary condition, we include both lowest KK mode \(n=0\) and the first excited KK mode \(n=1\). The gray region represent the square of the exponential factors: \eq{eq:tree-exp-sup} for boundary production and \eq{eq:pair-rate} (with \(\kappa=\mu_X\)) for pair production.}
    \label{fig:KK-Schwinger-bound}
\end{figure}

The semi-classical approximation provides a simple qualitative understanding of the phenomenon, yet it requires the limit \(\lambda_0L\gg1\) which is hard to achieve in practice. For finite \(L\), the situation is more complicated and a numerical approach is needed to determine the constraints. In the numerical setup, one computes the lowest eigenvalue \(E(\omega)\) of the ``Hamiltonian" in \eq{eq:SE}
for different \(\omega\). Boundary production can happen when \(E(0) < -\mu_X^2\), whereas pair production can happen when \(\max E(\omega) < -\mu_X^2\) for \(-\lambda_0<\omega<0\).
The results are summarized in \fig{fig:KK-Schwinger-bound} for both production mechanisms and different choices of boundary conditions.
Among different boundary conditions besides the zero mode of the double Neumann boundary conditions, the optimal choice is given by Neumann-Dirichlet boundary condition, where for the boundary production the breaking should happens at the Dirichlet boundary. In this case, for small \(\mu_X\), the boundary production requires \(\lambda_0 \gtrsim 0.675\Mc\) whereas the pair production requires \(\lambda_0 \gtrsim 2.33\Mc\).

\subsection{Signal estimates}

\label{subsec:chem-min-p}

In this section, we will estimate the signal size in the bispectrum from producing these KK excitations. We will focus on the boundary production mechanism as it requires much smaller \(\lambda_0/\Mc\) ratio and allows the signal to be generated at tree-level. The first step is to derive the interactions between the KK modes \(\chi_n\) and \(\phipt\). Consider the field redefinition:
\begin{align}
    X(x,x_5) &= e^{-iq_5(L-x_5)\gf{A}_5(x)} X_*(x,x_5).
\end{align}
The zeroth order equation of motion under the inflaton background is then given by
\begin{align}
    \p{-\partial_5^2 -\mathscr{D}_\mu^2 +m_X^2} X_* &= 0, &
    \mathscr{D}_\mu X_* &\coloneqq \p{\nabla_\mu -\frac{iQ}{f_2} \frac{L-x_5}{L}\nabla_\mu\phicl} X_*.
\end{align}
Substituting this into the action of \(X\) and neglect KK excitations in \(\gf{A}_M\), we find that
\begin{align}\label{eq:5D24D-interactions}
    S &\supset -\int \sqrt{-g} \dd[4]{x}  \bq{ \int_0^L \dd{x_5} \lbar{X}_*  \p{-\partial_5^2 -\mathscr{D}_\mu^2 +m_X^2} X_* +\frac{Q}{f_2} \nabla_\mu\phipt J^\mu
    +\frac{Q^2}{f_2^2} (\nabla_\mu\phipt)^2 \rho },
\end{align}
where
\begin{align}
    J_\mu &\coloneqq i\int_0^L \dd{x_5} \frac{L-x_5}{L} \p{\lbar X_* \mathscr{D}_\mu X_* -X_* \mathscr{D}_\mu \lbar{X}_*},&
    \rho &\coloneqq \int_0^L \dd{x_5} \p{\frac{L-x_5}{L}}^2 \abs{X_*}^2.
\end{align}
As a crude estimation, we will assume that \(\lambda_0 \sim\Mc\sim m_X\gg H\) are all comparable except in the exponential factors. After substituting in \(X_* \sim v_{XL} f_0 +\sum \chi_n f_n\) for the VEV and the KK excitations, \eq{eq:5D24D-interactions} introduces the following interactions:
\begin{align}
    \Mc^{3/2}\bar v_{XL}\exp\p{-\frac{\pi^2 \mu_X^2}{4\lambda_0 \Mc}} \cdot \p{\frac{Q\phipt}{f_2}}^k \chi_n,&&
    \Mc^2\cdot\p{\frac{Q\phipt}{f_2}}^k \bar\chi_n \chi_m,&& k &= 1,2.
\end{align}
Here we have substituted all the derivatives by \(\Mc\) since the energy and momentum scale at the on-shell exchange are given by \(\lambda_0,m_X\sim \Mc\).

\begin{figure}
    \centering
    \begin{tabular}{c@{\hspace{5em}}c}
        \includegraphics{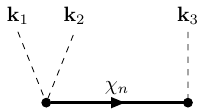} &
        \includegraphics{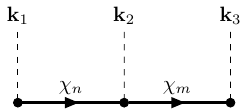}\\
        \hspace{0.9em} single exchange & double exchange
    \end{tabular} 
    \caption{Two types of tree-level Feynman diagrams contributing to the 3-point correlator. The diagrams are labeled based on the number of \(\chi\) propagators. Note that for each diagrams one must sum over all permutations of the external momenta to obtain the full 3-point correlator.}
    \label{fig:5D-Feynman}
\end{figure}

From these interactions, one can construct two types of tree-level diagrams contributing to the 3-point correlator as shown in \fig{fig:5D-Feynman}.
The signal size coming from the single-exchange diagram can be estimated as (\(H=1\))
\begin{align}\label{eq:SE-NG-estimation}
    \Fosc^{\text{SE}} &\sim \dphicl \p{\frac{Q}{f_2}}^3 \bq{\Mc^{3/2} \abs{v_{XL}} \exp\p{-\frac{\pi^2 \mu_X^2}{4\lambda_0 \Mc}} }^2\n
    &\peq \times \int a^4(\eta_R) d(k,\eta_R) u_n(k,\eta_R) \dd{\eta_R} \int a^4(\eta_L)  \bar d^2(k,\eta_L) \bar u_n(k,\eta_L) \dd{\eta_L},
\end{align}
where \(a(\eta) \coloneqq 1/(-H\eta)\) is the scale factor,
\begin{align}
    u_n(k,\eta) \sim \frac{a^{-3/2}(\eta)}{\sqrt{2\omega_n(k,\eta)}} \exp\bq{-i\int a(\eta) \omega_n(-k\eta) \dd{\eta}}
\end{align}
is the 4D wave function under the adiabatic approximation and
\(d(k,\eta) \sim a^{-1}(\eta) e^{-ik\eta}\) is the bulk-boundary propagator of the inflaton before exiting the horizon. In the single-exchange diagram, both vertices in \fig{fig:5D-Feynman} can be simultaneously on-shell, so both time-integrals in \eq{eq:SE-NG-estimation} can be estimated using the stationary phase approximation:
\begin{align}
    \int a^4(\eta) d^n(k,\eta) u_n(k,\eta) \dd{\eta} &\sim \frac{a^{-n +3/2}(\eta_*)}{\omega_n(-k\eta_*)} \sim k^{-n +3/2} \omega_n^{n-5/2}(-k\eta_*).
\end{align}
With \(\omega_n(-k\eta_*)\sim\Mc\), we thus find
\begin{align}
    \Fosc^{\text{SE}} &\sim \frac{\dphicl}{\Mc^2} \p{\frac{Q}{f_2}}^3 \bq{\Mc^{3/2} \abs{v_{XL}} \exp\p{-\frac{\pi^2 \mu_X^2}{4\lambda_0 \Mc}} }^2\n
    &\sim \frac{1}{\epsilon_H} \frac{\Mc\abs{v_{XL}}^2}{V_0} \p{\frac{\Mc}{H}}^3 \exp\p{-\frac{\pi^2 \mu_X^2}{2\lambda_0 \Mc}}.
\end{align}
On the other hand, for the double-exchange diagram in \fig{fig:5D-Feynman}, one can use the same method to estimate the signal size. There are two cases:
\begin{enumerate}
    \item When \(n=m\), the middle vertex is always off-shell, so the leading contribution comes from the coincident limit where the middle vertex coincides with one of the other two vertices, and we have \(\Fosc^{\text{DE}} \sim \Fosc^{\text{SE}}\).

    \item When \(n\ne m\), the middle vertex can also be on-shell and contribute an extra factor
    \begin{equation}
        \Mc^2 \int a^4(\eta) e^{-ik\eta} u_n(k,\eta) \bar u_m(k,\eta) \dd{\eta} \sim \sqrt{\Mc}.
    \end{equation}
    In this case, we have \(\Fosc^{\text{DE}} \sim \sqrt{\Mc/H} \cdot \Fosc^{\text{SE}}\).
\end{enumerate}
Adding both contributions together, we expect
\begin{align}\label{eq:SE-NG-estimation-2}
    \Fosc &\sim \frac{1}{\epsilon_H} \frac{\Mc\abs{v_{XL}}^2}{V_0} \p{\frac{\Mc}{H}}^{3 +\delta/2} \exp\p{-\frac{\pi^2 \mu_X^2}{2\lambda_0 \Mc}},
\end{align}
where \(\delta = 0\) if only the lightest KK mode can be produced; \(\delta = 1\) if at least one excited KK mode can also be produced.

In order to constrain the signal size \eq{eq:SE-NG-estimation-2}, we need to find constraints on the VEV \(v_{XL}\). Paralleling the consideration of \eq{eq:potential-bound-s}, we have
\begin{align}\label{eq:SE-NG-vmax}
    \abs{v_{XL}}^2 &< \frac{\LambdafD^3}{16\pi^2}.
\end{align}
For an inflaton potential \(V_0\) saturating \eq{eq:potential-bound-w}, this points to
\begin{align}\label{eq:SE-NG-estimation-3}
    \Fosc &\sim \frac{1}{N\epsilon_H} \p{\frac{\Mc}{H}}^{3 +\delta/2} \exp\p{-\frac{\pi^2 \mu_X^2}{2\lambda_0 \Mc}}.
\end{align}
At the tri-axion benchmark \eq{eq:KK-benchmark}, this na\"ively corresponds to \(\Fosc = \order\p{10^6}\) if the exponential is \(\order(1)\). However, for large \(\Fosc\), one must also consider the back-reaction to the power spectrum as in \eq{eq:power-spectrum-constraint}. Calculating this back-reaction using the same estimate as \eq{eq:SE-NG-estimation-2} shows
\begin{align}
    \Fosc &\sim \frac{\delta\Ps}{\Ps} \p{\frac{\Mc}{H}}^{2 +\delta/2}.
\end{align}
To avoid strong back-reactions, the maximal non-Gaussianity is thus constrained to be \(\Fosc\le \order\p{10^3}\) for \(\delta\Ps<0.1\Ps_0\). Since large suppression from the exponential requires large \(\lambda/\Mc\) ratio as shown in \fig{fig:KK-Schwinger-bound}, the \(\delta\Ps/\Ps\) constraint is more likely to be achieved with \(\abs{v_{XL}}^2\) being \(\order\p{10^{-3}-10^{-2}}\) smaller than its theoretical maximal value in \eq{eq:SE-NG-vmax}.

There is another concern with having new charged particles around \(\Mc\) from the quality problem as they may introduce dangerous new corrections to the inflaton potential. To see this, note that if \(X\) also acquires a VEV \(v_{X0}\) on the \(x_5=0\) brane, there will be a new contribution to the inflaton potential from mixing the two VEVs:
\begin{equation}
    \delta V \sim \Mc \abs{v_{X0} \bar v_{XL}} \exp\p{-\frac{\pi^2 \mu_X^2}{4\lambda_0 \Mc}} \cos\p{\frac{Q\phi_2}{f_2} +\delta}.
\end{equation}
Here the tunneling factor should also be given by \eq{eq:tree-exp-sup} instead of the na\"ive Yukawa suppression \(e^{-m_X L}\) in \eq{eq:potential-height}.\footnote{This issue does not occur for the axion potential from \(\gf{H}_{1,2}\) because their chemical potentials are small.} For \(\abs{v_{XL}}\sim \abs{v_{X0}}\), this implies
\begin{align}
    \Fosc &\sim \frac{1}{\epsilon_H} \frac{\delta V_0}{V_0} \p{\frac{\Mc}{H}}^{3 +\delta/2} \exp\p{-\frac{\pi^2 \mu_X^2}{4\lambda_0 \Mc}}.
\end{align}
Compared with \eq{eq:mod-constraint}, we see that the size of \(\Fosc\) can still be as large as \(\order\p{10^3}\) for \(\order(1)\) exponential.
In particular, for nonzero \(\abs{v_{X0}}\), the \(\chi\) particle can be produced and annihilated on both sides of the boundary and thus generating scale-invariance-breaking cosmological collider signal without Boltzmann suppression as discussed in \cite{Flauger:2016idt,Chen:2022vzh}. From the same estimation, the size of this signal is roughly smaller by a factor
\begin{equation}
    \frac{\abs{v_{X0}v_{XL}}}{\abs{v_{X0}}^2 +\abs{v_{XL}}^2}
\end{equation}
compared to the scale-invariant signal as it requires symmetry breaking on both boundaries. They are still of interest as they serve as strong evidence of the multi-axion origin of the inflaton potential.
Note however that naturalness does not imply \(\abs{v_{X0}}\sim \abs{v_{XL}}\) as one can simply gauge the \(\mathrm{U}(1)\) symmetry of \(X\) at \(x_5=0\) to forbid \(v_{X0}\). For example, if \(X\) is charged under a new \(\mathrm{U}(1)\) gauge boson \(\gf{A}^{(X)}_M\) with a mixed boundary condition:
\begin{align}
    \eval{\gf{F}^{(X)}_{\mu5}}_{x_5=0} &= 0,&
    \eval{\gf{A}^{(X)}_\mu}_{x_5=L} &= 0,
\end{align}
\(v_{X0}\) will be forbidden by the \(\mathrm{U}(1)_X\) gauge symmetry whereas \(v_{XL}\) will still be allowed.

\section{Constraining Chemical Potentials from Non-minimal Interactions}

\label{sec:chem-nonmin-p}

In this section, we will work out the maximal chemical potentials and signal sizes for chemical potentials from non-minimal interactions in \Sec{subsec:nonmin} by applying the constraints in \Sec{subsec:constraints}.
Recall that the size of chemical potentials from non-minimal interactions are given by \eq{eq:chem-pot-non-minimal}. The potential height constraints \eqs{eq:inflaton-potential}{eq:potential-bound-w} then imply
\begin{equation}\label{eq:chem-pot-non-minimal-bound}
    \lambda \lesssim \frac{2\dphicl}{\sqrt{\Mpl H}} \p{\frac{4N}{3\pi}}^{1/4} \p{\frac{\Mc}{\LambdafD}}^{\Delta -3/4} \approx 29N^{1/4}\p{\frac{\Mc}{\LambdafD}}^{\Delta -3/4} H.
\end{equation}
Consider the following tri-axion benchmark with \(f_1=f_2=f_3=f\), \(N_1=N_2=N\) and the minimal \(\LambdafD/\Mc = 4.5\) suggested by the quality problem in \eq{eq:quality-bound}:
\begin{align}\label{eq:KK-benchmark-2}
    N &= 25,&
    \Mc &= 160H,&
    \LambdafD &= 720H,&
    f &= 640H,
\end{align}
the maximal chemical potential is then
\begin{align}
    \lambda_{3/2} &\lesssim 21H\quad\text{for }\Delta = 3/2,&
    \lambda_{5/2} &\lesssim 4.7H\quad\text{for }\Delta = 5/2,
\end{align}
whereas chemical potentials from higher \(\Delta\) are too small to be relevant.
Note that the same constraints also imply
\begin{equation}
    \frac{\lambda}{\Mc} \lesssim \sqrt{\frac{8N\epsilon_H}{3\pi^2}} \p{\frac{\Mc}{\LambdafD}}^{\Delta-3/2} \lesssim 0.034\sqrt{N} \approx \begin{dcases}
        0.54, & N=249\\
        0.22, & N=39.
    \end{dcases}
\end{equation}
where \(N=249,\ 39\) are the maximal possible values for the bi-axion and the tri-axion models in \tab{tab:Mc-min}, respectively. Therefore, KK excitations are very unlikely to be produced with only chemical potentials from non-minimal interactions, and we are left with the lightest KK modes and brane-localized modes if they have \(m\ll\Mc\). 

\begin{table}
    \centering
    \begin{tabular}{lll}\toprule
         5D EFT & 4D EFT & comments\\\midrule
         \(\gf{F}_{MN} \bar\Psi \bq{ \gamma^M, \gamma^N } \Psi\) & \(\nabla_\mu\phi\:\bar\psi_n \gamma^\mu\gamma_5 \psi_n\) & bulk fermion \(\Psi\)\\
         \(\epsilon^{MNPQR} \gf{A}_M \gf{G}_{NP} \gf{G}_{QR}\) & \(\epsilon^{\mu\nu\rho\sigma} \phi\:\! \lbar{G}^{(n)}_{\mu\nu} G^{(n)}_{\rho\sigma}\) & bulk vector \(\gf{B}\) with field strength \(\gf{G}\)\\
         \(X \gf{F}_{MN} \gf{G}^{MN}\) & \(\chi_0 \nabla_\mu \phi \nabla^\mu b_5\) & bulk scalar \(X\) + bulk vector \(\gf{B}\)\\
         & & with Dirichlet boundary conditions (axion)\\\bottomrule
    \end{tabular}
    \caption{The dimension-\(6\frac{1}{2}\) (\(\Delta=3/2\)) bulk operators that can give rise to chemical potentials in 4D.}
    \label{tab:D=3/2-operators}
\end{table}

At the lowest \(\Delta = 3/2\) order, there are only five operators that can give rise to chemical potential operators in the 4D EFT. These are three bulk operators shown in \tab{tab:D=3/2-operators} and the following two brane operators for brane-localized scalars and fermions:
\begin{align}\label{eq:nm-boundary}
    \gf{F}_{\mu5} J^\mu,
    && J^\mu &= \begin{dcases}
        i\p{\bar\chi\nabla^\mu\chi -\chi\nabla^\mu\bar\chi},& \text{for scalar \(\chi\)}\\
        \psi\sigma^\mu\psi^\dagger,& \text{for fermion \(\psi\)}.
    \end{dcases}
\end{align}
Besides the two example models in \Sec{subsec:constraints}, these also included the 5D generalization of the fermionic chemical potential discussed in \cite{Chen:2018xck} for both brane and bulk fermions,
as well as a new possibility of a chemical potential between a bulk scalar and an axion from some bulk gauge field, similar to the scalar model in \Cite{Bodas:2020yho,Bodas:2025wuk}.

We now briefly comment on the signal size generated by the two example models in \Sec{subsec:nonmin}. For the brane scalar model given by \eq{eq:brane-scalar-model}, the leading-order cosmological collider signals are generated at tree-level. For the brane scalar model, the signal size in the bispectrum is straightforwardly given by \eq{eq:f-arushi-model} or \eqref{eq:f-arushi-model-2}. With \(\lambda=10H\) for example, we find \(\Fosc\le \order\p{10}\) for \(\delta\Ps<0.1\Ps_0\), much lower than the possible signal size from \eq{eq:SE-NG-estimation-2} but still observable by ongoing/upcoming experiments.
Meanwhile, the signal from the gauge boson model \eq{eq:KK-gauge-model} comes from one-loop process and is much harder to estimate.
The 4D EFT \eq{eq:KK-gauge-4D} has been studied in \Cite{Wang:2020ioa}. It is found that the loop-suppression can be overcome by the exponential enhancement \eq{eq:helicity-exp-enhance} as well as some phase-space enhancement. While the estimates in \Cite{Wang:2020ioa} suggest that an observable signal can be generated within theoretical control, there remains significant uncertainties in the estimation, warranting further study.

\section{Discussion}

\label{sec:discussion}

Higher-dimensional gauge theory provides a natural platform for realizing robust high-scale inflation.
In this paper, we have shown that the minimal gauge interactions in this setting automatically give rise to chemical potentials for bulk charges via an extra-dimensional Schwinger effect. We found realistic models and benchmarks where KK excitations can be produced and observable cosmological collider signals are generated. We also found chemical potentials from non-minimal gauge interactions and derived constraints on their mass reach and signal strength.

\begin{table}
    \centering
    \begin{tabular}{lll}\toprule
         5D SUSY & 4D SUSY & 4D QFT\\\midrule
         \multirow{2}[1]{*}{\makecell{\(\mathscr{V}\mathscr{V}'\mathscr{V}'\)\\(Chern-Simons)}} & \(\int\dd[2]{\theta} \Phi \gf{W}'^\alpha \gf{W}'_\alpha\) & \(\epsilon^{\mu\nu\rho\sigma} \phi F'_{\mu\nu} F'_{\rho\sigma},\ 
         \nabla_\mu \phi \lambda' \sigma^\mu\lambda'^\dagger\)\\
         & \(\int \dd[4]{\theta} (\Phi +\bar\Phi) (\Phi' +\bar\Phi')^2\) & \(\sigma' \nabla^\mu \phi \nabla_\mu \phi'\)\\\midrule
         - & \(\int \dd[4]{\theta} \p{\Phi +\bar\Phi -\sqrt{2}\partial_5 \mathcal{V}} \lbar XX\) & \(\nabla_\mu\phi J^\mu\)\\\bottomrule
    \end{tabular}
    \caption{All non-minimal interactions in \tab{tab:D=3/2-operators} can arise from the super-Chern-Simons interactions in 5D SUSY for bulk vector multiplets \cite{Arkani-Hamed:2001vvu}. The two brane chemical potential in \eq{eq:nm-boundary} can also arise from a brane K\"ahler potential for brane-localized chiral multiplets \(X\) \cite{Bodas:2025wuk} Under the 4D subalgebra of the 5D SUSY, a 5D vector supermultiplet \(\mathscr{V}\) decomposes into a 4D vector multiplet \(\mathcal{V}\) and a 4D chiral multiplet \(\Phi\) (with \(x_5\) as an invariant label). \(\mathcal{V}\) consists of a vector \(\gf{A}_\mu\) and a gaugino \(\lambda\); \(\Phi\) consists of a fermion \(\psi\) and a complex scalar \(\sigma +i\phi\) with \(\phi\propto\gf{A}_5\).}
    \label{tab:non-min-SUSY}
\end{table}

We realized the inflaton potential at tree-level, generated through the exchange of light charged bulk scalars. To keep these scalars naturally light compared to the UV cutoff \(\LambdafD\), it is attractive to embed this model into a supersymmetric (SUSY) theory where these charged scalars become the scalar components of some hypermultiplets. It is also highly plausible that SUSY is relevant at this extremely high energy, motivated by the electroweak hierarchy problem and grand unification or simply as a remnant of string theory.
The chemical potentials arising from minimal gauge interactions are straightforwardly generalized with supersymmetric gauge interactions.
The other bulk chemical potential interactions from non-minimal interactions we have discussed are captured by the 5D Chern-Simons terms and related interactions by 5D SUSY \cite{Arkani-Hamed:2001vvu}.
The two types of brane-localized chemical potentials can arise from a brane-localized K\"ahler potential for a brane-localized chiral multiplet \cite{Bodas:2025wuk}. For the various operator structures for the non-minimal interactions, see \tab{tab:non-min-SUSY}.

We have been focusing on the higher-dimensional multi-axion model as a solution to the trans-Planckian problem and the quality problem of high-scale inflation. 
Besides extra dimensions, the quality problem can also be solved purely in 4D if these axions are composite particles from some strong dynamic. If the strong dynamic is approximately conformal over a large hierarchy, the physics can again be understood through their higher-dimensional holographic duals, except that the extra dimension now must be \emph{warped}.
Under the duality, a bulk gauge boson in the higher-dimensional theory with Dirichlet boundary conditions is dual to an approximate global symmetry of the composite sector, spontaneously broken at the compactification scale,\footnote{The Dirichlet boundary condition on the IR brane comes from a brane-localized Higgs in the \(g_5 v_{\text{Higgs}}\to\infty\) limit.} where the axion is the PNGB from this symmetry breaking.
The axion potential is then generated by an explicit irrelevant breaking of the symmetry, dual to the bulk charged scalar \eq{eq:tree-A5-potential} with inhomogeneous boundary conditions in the higher-dimensional theory.
Working in the limit \(k/H \gg e^{kL} \gg 1\) where \(k\) is the 5D curvature, and repeating the analysis in \Sec{subsec:orbifolding}, it is straightforward to find that most of our discussion is still applicable up to \(\order(1)\) factors provided one makes the following replacement:
\begin{align}
    \Mc &= \frac{\pi}{L} \to \Mc = ke^{-kL},&
    \LambdafD &\to \LambdaIR \coloneqq \LambdafD e^{-kL},&
    f &= \frac{1}{g_5\sqrt{L}} \to f = \frac{\sqrt{2k}}{g_5} e^{-kL}.
\end{align}
There is one important exception coming from the new WGC constraint:
\begin{equation}
    f\lesssim \p{\frac{\Mc}{\LambdafD}} \Mpl = \p{\frac{\Mc}{\LambdaIR}} \Mpl e^{-kL},
\end{equation}
which becomes tighter as the warp factor \(e^{kL}\) increases. Consequently, production of KK excitations seems to require either a modest warp factor or including more axions. It remains interesting to see whether a compactification scale within the reach of chemical potentials is consistent with a stronger warping in multi-axion extensions.

Finally, there exist other mechanisms that can solve the field range problem \eq{eq:trans-Lambda} besides the aligned axion mechanism we considered. For example, ``axion monodromy" invokes multi-valued axion potentials that effectively extends the field range beyond the axion decay constant \cite{Silverstein:2008sg,McAllister:2008hb}. There is a straightforward generalization of our construction to the case of axion monodromy, but it remains to be seen whether particle production from chemical potentials is parametrically consistent with all the theoretical constraints in these models.

\acknowledgments

This work was supported by NSF grant PHY-2514660 and by the Maryland Center for Fundamental Physics.

\appendix

\bibliographystyle{JHEP}
\bibliography{main.bib}

\end{document}